\documentclass[fleqn,usenatbib]{mnras}

\usepackage{newtxtext,newtxmath}

\usepackage[T1]{fontenc}
\usepackage[usenames,dvipsnames,svgnames,table]{xcolor}
\usepackage{subcaption}
\usepackage{multirow}
\usepackage{enumitem}
\usepackage[flushleft]{threeparttable}

\DeclareRobustCommand{\VAN}[3]{#2}
\let\VANthebibliography\thebibliography
\def\thebibliography{\DeclareRobustCommand{\VAN}[3]{##3}\VANthebibliography}

\newsavebox{\measurebox}

\usepackage{graphicx}	
\usepackage{amsmath}	
\usepackage{scalerel,tikz}
\usetikzlibrary{svg.path}
\definecolor{orcidlogocol}{HTML}{A6CE39}
\tikzset{orcidlogo/.pic={
		\fill[orcidlogocol] svg{M256,128c0,70.7-57.3,128-128,128C57.3,256,0,198.7,0,128C0,57.3,57.3,0,128,0C198.7,0,256,57.3,256,128z};
		\fill[white] svg{M86.3,186.2H70.9V79.1h15.4v48.4V186.2z}
		svg{M108.9,79.1h41.6c39.6,0,57,28.3,57,53.6c0,27.5-21.5,53.6-56.8,53.6h-41.8V79.1z M124.3,172.4h24.5c34.9,0,42.9-26.5,42.9-39.7c0-21.5-13.7-39.7-43.7-39.7h-23.7V172.4z}
		svg{M88.7,56.8c0,5.5-4.5,10.1-10.1,10.1c-5.6,0-10.1-4.6-10.1-10.1c0-5.6,4.5-10.1,10.1-10.1C84.2,46.7,88.7,51.3,88.7,56.8z};
}}
\newcommand\orcidicon[1]{\href{https://orcid.org/#1}{\mbox{\scalerel*{
				\begin{tikzpicture}[yscale=-1,transform shape]
					\pic{orcidlogo};
				\end{tikzpicture}
			}{|}}}}

\usepackage{hyperref}

\usepackage[table]{xcolor}

\definecolor{fiducialcol}{HTML}{D90C10}
\definecolor{agncol}{HTML}{5DA011}
\definecolor{cgmcol}{HTML}{055FA8}
\definecolor{ssfrcol}{HTML}{611C6B}
\definecolor{sfrcol}{HTML}{8A4F14}
\definecolor{mstarcol}{HTML}{E17334}
\definecolor{sfdisccol}{HTML}{BA129B}
\providecommand{\modelcolorbox}{}
\renewcommand{\modelcolorbox}[1]{%
	\begingroup
	\setlength{\fboxsep}{0pt}%
	\colorbox{#1}{\rule{1.4em}{0pt}\rule{0pt}{0.8em}}%
	\endgroup
}

\defcitealias{Qin2025PASA...42...49Q}{Q25}

\title[forest inference with SAM]{Robust Ly$\alpha$ forest constraints on reionization from semi-analytic galaxy formation models}

\author[Qin et al.]{Yuxiang Qin~\orcidicon{0000-0002-4314-1810}$^{1}$\thanks{E-mail: Yuxiang.L.Qin@gmail.com} and J. Stuart B. Wyithe~\orcidicon{0000-0001-7956-9758}$^{1}$
\\
$^{1}$Research School of Astronomy and Astrophysics, Australian National University, Canberra, ACT 2611, Australia
}

\date{}

\pubyear{2026}

\begin{document}
\label{firstpage}
\pagerange{\pageref{firstpage}--\pageref{lastpage}}
\maketitle

\begin{abstract}
	The high-redshift Ly$\alpha$ forest provides one of the most sensitive probes of the final stages of reionization, but its interpretation can depend on how the ionizing sources are modelled. We test the robustness of Ly$\alpha$ forest-inferred reionization histories using \textsc{Meraxes}, a semi-analytic model of galaxy formation and reionization applied to a cosmological volume of $(210h^{-1}{\rm cMpc})^3$ with merger trees augmented down to the atomic-cooling threshold. We construct a suite of source models in which the galaxy escape fraction, $f_{\rm esc}^{\rm gal}$, is parameterized as a function of halo mass, stellar mass, star formation rate, specific star formation rate, or star-forming disc surface density. We also consider models in which accreting blackholes contribute ionizing photons, or dense circumgalactic gas further attenuates the escape of ionizing photons. Each model is calibrated to the observed distribution of Ly$\alpha$ effective optical depths from XQR-30+.
	These prescriptions redistribute the escaping photon budget across halo mass and UV luminosity, introduce scatter in $f_{\rm esc}^{\rm gal}$ at fixed halo mass, and allow modest variation in the early evolution of reionization. Nevertheless, all viable calibrated models recover a late and extended EoR, with the midpoint occurring at $z\sim6.2$--6.9 and the final stages at $z\sim5.36$--5.57. Our results indicate that the late reionization history inferred from Ly$\alpha$ forest is robust to physically motivated scatter in the galaxy--halo connection, modest AGN contributions, and phenomenological CGM attenuation.
\end{abstract}




\section{Introduction}

The Epoch of Reionization (EoR) marks the final major phase transition of baryonic matter in the Universe, during which the intergalactic medium (IGM) evolved from being predominantly neutral to highly ionized \citep{Barkana2000, Furlanetto2006PhR...433..181F,Mesinger2019cosm.book.....M}. Establishing when and how this transition occurred is fundamental, because the timing, duration, and morphology of reionization encode the abundance and nature of the first luminous sources, their ionizing photon production efficiencies, and the impact of radiative feedback on early galaxy formation \citep{Dayal2018PhR...780....1D,Robertson2022ARA&A..60..121R}. Over the past decade, increasingly precise measurements of the cosmic microwave background (CMB) optical depth (e.g., \citealt{Planck2020A&A...641A...6P,Reichardt2021ApJ...908..199R}), high-redshift galaxy luminosity functions (LFs; e.g., \citealt{Bouwens2023MNRAS.523.1036B,Harikane2023ApJS..265....5H}), quasar damping wings (e.g., \citealt{Greig2024MNRAS.530.3208G,Zhu2024MNRAS.533L..49Z}), Ly$\alpha$ emitter (LAE) statistics (e.g., \citealt{Tang2024ApJ...975..208T,Umeda2025ApJS..277...37U}), and the Ly$\alpha$ forest (e.g., \citealt{Fan2006AJ....132..117F,Bosman2022MNRAS.514...55B}) have all made substantial progress in constraining the reionization history. 

Among the available probes, the high-redshift Ly$\alpha$ forest provides a uniquely sensitive measurement of the ionization and thermal state of the late- and post-reionization IGM. The effective optical depth, $\tau_{\rm eff}$, measured over large segments of quasar spectra, responds not only to the mean photoionization rate, but also to spatial fluctuations in the ultraviolet background (UVB; \citealt{Cain2021ApJ...917L..37C,Davies2016MNRAS.460.1328D}), the temperature-density relation of the IGM \citep{DAloisio2015ApJ...813L..38D,Keating2018MNRAS.477.5501K}, the underlying density field \citep{Bolton2017MNRAS.464..897B,Walther2019ApJ...872...13W}, and residual neutral islands left over from a patchy and incomplete reionization process \citep{Qin2021MNRAS.506.2390Q,Cain2024MNRAS.531.1951C}. Observations at $z \gtrsim 5$ reveal large sightline-to-sightline scatter in Ly$\alpha$ opacity \citep{Becker2018ApJ...863...92B,Zhu2021ApJ...923..223Z}, including extended \citet{gunn1965density} troughs that are difficult to reproduce with a spatially uniform ionizing background alone. These opacity fluctuations therefore provide a powerful probe of the late stages of reionization, complementing global constraints from the CMB and galaxy surveys.

\begin{figure}
	\centering
	\includegraphics[width=\columnwidth]{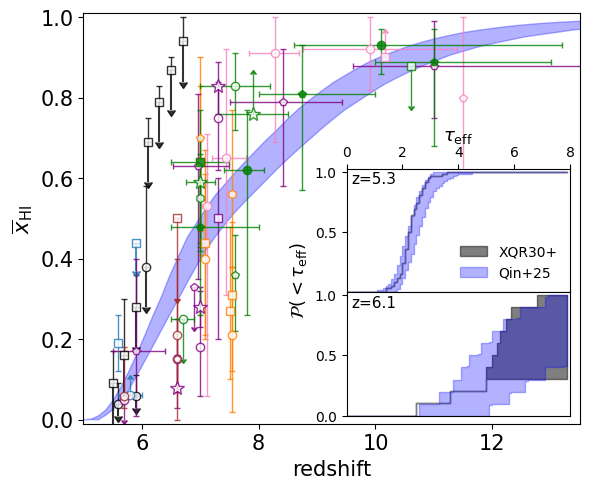}
	\caption{Summary of the Ly$\alpha$ forest constraints on reionization history inferred by \citetalias{Qin2025PASA...42...49Q}, shown here to motivate the present work. 
	The main panel shows the reconstructed evolution of the volume-averaged neutral fraction, where the blue shaded region indicates the 2.5--97.5\% of the Constant-$f_{\rm esc}$ posterior from \citetalias{Qin2025PASA...42...49Q}. For clarity, the observational labels are omitted here and are given in Fig.~\ref{fig:xH_mvir}. 
	The inset panels show two example cumulative distributions of Ly$\alpha$ effective optical depth, $\tau_{\rm eff}$, at $z=5.3$ and 6.1, comparing \citetalias{Qin2025PASA...42...49Q} with the XQR-30+ measurements \citep{Bosman2022MNRAS.514...55B}.
	This figure illustrates that the Ly$\alpha$ forest implies a tight late reionization history, motivating the main goal of this paper, which is to test how robust that inference remains when more flexible galaxy-formation models are adopted.}
	\label{fig:Qin25}
\end{figure}

In previous work, we demonstrated that forward-modelling the high-redshift Ly$\alpha$ forest can yield very tight constraints on the reionization history (\citealt{Qin2021MNRAS.506.2390Q}; \citealt{Qin2025PASA...42...49Q}, hereafter \citetalias{Qin2025PASA...42...49Q}). In particular, \citetalias{Qin2025PASA...42...49Q} showed that the observed distribution of Ly$\alpha$ effective optical depths strongly restricts the timing of the final overlap phase of reionization when combined with existing galaxy and CMB constraints. As illustrated in Fig.~\ref{fig:Qin25}, this analysis favours a relatively tight late-time ionization history and successfully reproduces the observed Ly$\alpha$ opacity distributions at both moderate and high effective optical depths. This constraining power arises because the Ly$\alpha$ forest is sensitive to residual neutral hydrogen fractions as low as $\sim10^{-4}$, so even modest changes in the timing, topology, or spatial uniformity of reionization can leave measurable imprints on the opacity distribution.

At the same time, such tight constraints invite an important question: {\it how robust are they to the modelling of the ionizing source population?} A key limitation of many existing EoR inference frameworks is that, when sources are actually modelled explicitly rather than absorbed into effective IGM parameters (e.g., \citealt{Gaikwad2023MNRAS.525.4093G,Davies2024ApJ...965..134D,Cain2025PASA...42..107C} for the latter), they are often described using relatively simple scaling relations between galaxy properties and the density field or halo mass (e.g., \citealt{Choudhury2018MNRAS.481.3821C,Park2019MNRAS.484..933P}). For example, the stellar-to-halo mass relation, ionizing escape fraction, and star-formation suppression threshold are frequently parameterized as smooth functions of halo mass, with a small number of global parameters controlling the ionizing photon budget. Such models are computationally efficient and well suited to Bayesian inference, but they necessarily compress the complexity of galaxy formation into deterministic or near-deterministic mappings.

However, high-redshift galaxies are expected to exhibit substantial stochasticity in their stellar masses, star-formation histories, gas accretion rates, feedback strengths, and escape fractions \citep{Paardekooper2015MNRAS.451.2544P,Ma2020MNRAS.498.2001M,Yeh2023MNRAS.520.2757Y}. Similarly, photoheating feedback, governed by the local UVB intensity and ionization history, is unlikely to be captured fully by a single constant critical halo mass below which star formation is suppressed \citep{Sobacchi2013MNRAS.432L..51S,Ocvirk2020MNRAS.496.4087O}. These simplifications may artificially restrict the range of reionization histories and ionizing-background fluctuations that can be produced, potentially causing Ly$\alpha$ forest constraints to appear tighter than they would be in a more physically flexible galaxy-formation model. In this sense, a more detailed galaxy-formation model provides an important estimate of the systematic uncertainty on the statistical EoR constraints inferred from the Ly$\alpha$ forest.

In this work, we therefore revisit Ly$\alpha$ forest constraints on the EoR using \textsc{Meraxes}, a semi-analytic model of galaxy formation and reionization \citep{Mutch2016,Qin2023MNRAS.526.1324Q}. Unlike simple halo-based emissivity prescriptions, \textsc{Meraxes} follows {\it N}-body merger trees and the assembly histories of galaxies, including gas accretion and cooling, star formation, supermassive blackhole growth, feedback from supernovae and active galactic nuclei, and the impact of reionization feedback on galaxy formation. We explore a suite of models in which the ionizing escape fraction is parameterized not only as a function of halo mass, but also as a function of galaxy properties such as stellar mass, star formation rate, specific star formation rate, and star-forming disc surface density. We also consider two extensions to the fiducial halo-mass-dependent model where (1) active galactic nuclei (AGN) contribute to the ionizing photon budget, motivated by recent {\it JWST} discoveries of faint AGN \citep{Labbe2023Natur.616..266L,Matthee2024ApJ...963..129M}; and (2) the escape of ionizing photons is further suppressed by dense structures in the circumgalactic medium (CGM) whose opacity increases with the local clumping factor. By comparing these models after calibration to the XQR-30+ Ly$\alpha$ forest measurements \citep{Bosman2022MNRAS.514...55B}, we test whether stochasticity and added flexibility in the source population substantially broaden the range of reionization histories compatible with the data, or whether the tight late-time EoR history inferred by \citetalias{Qin2025PASA...42...49Q} remains robust.

This paper is organized as follows. In Sec.~\ref{sec:model}, we describe the numerical framework used to predict reionization and galaxy-formation observables, including the \textsc{Meraxes} galaxy-formation model, the UV ionizing escape-fraction prescriptions, the ionization-field calculation, and the Ly$\alpha$ forest optical depth forward modelling.
Sec.~\ref{sec:fesc} presents our suite of Ly$\alpha$ forest-calibrated escape fraction models and examines how they redistribute the escaping ionizing photon budget across halo mass and UV luminosity.
In Sec.~\ref{sec:robust_eor}, we test the robustness of the resulting reionization histories against variations in the source model. Sec.~\ref{sec:discussion} presents a summary discussion before we conclude in Sec.~\ref{sec:conclusion}. Throughout this work, we assume the \citet{Kroupa2001MNRAS.322..231K} initial mass function (IMF) where appropriate, and adopt a flat $\Lambda$CDM cosmology with $(\Omega_{\rm m},\Omega_{\rm b},\Omega_{\Lambda},h,\sigma_8,n_{\rm s})=(0.312,0.0490,0.688,0.675,0.815,0.968)$ from \citet{Planck2020A&A...641A...6P}.

\section{Reionization and galaxy-formation observables from numerical simulations}\label{sec:model}

In this section, we describe the theoretical framework used to connect high-redshift galaxy formation, reionization, and Ly$\alpha$ forest transmission. We combine the \textsc{Meraxes} semi-analytic model of galaxy formation with a large-volume $N$-body simulation to predict the spatial distribution of galaxies and their ionizing emissivity during the EoR. These outputs are coupled to an inhomogeneous reionization calculation and a Ly$\alpha$ forest forward-modelling module, which converts the simulated IGM fields into distributions of effective optical depth. The Ly$\alpha$ forest calculation also includes a correction for unresolved small-scale opacity fluctuations calibrated against high-resolution hydrodynamical simulations. Together, this framework allows us to revisit Ly$\alpha$ forest constraints on the reionization history using a galaxy-formation model with substantially more flexibilities in the connection between galaxies, dark matter halos, and ionizing sources.

\subsection{The \textsc{Meraxes} galaxy-formation model}
\begin{figure*}
	\centering
	\includegraphics[width=\textwidth]{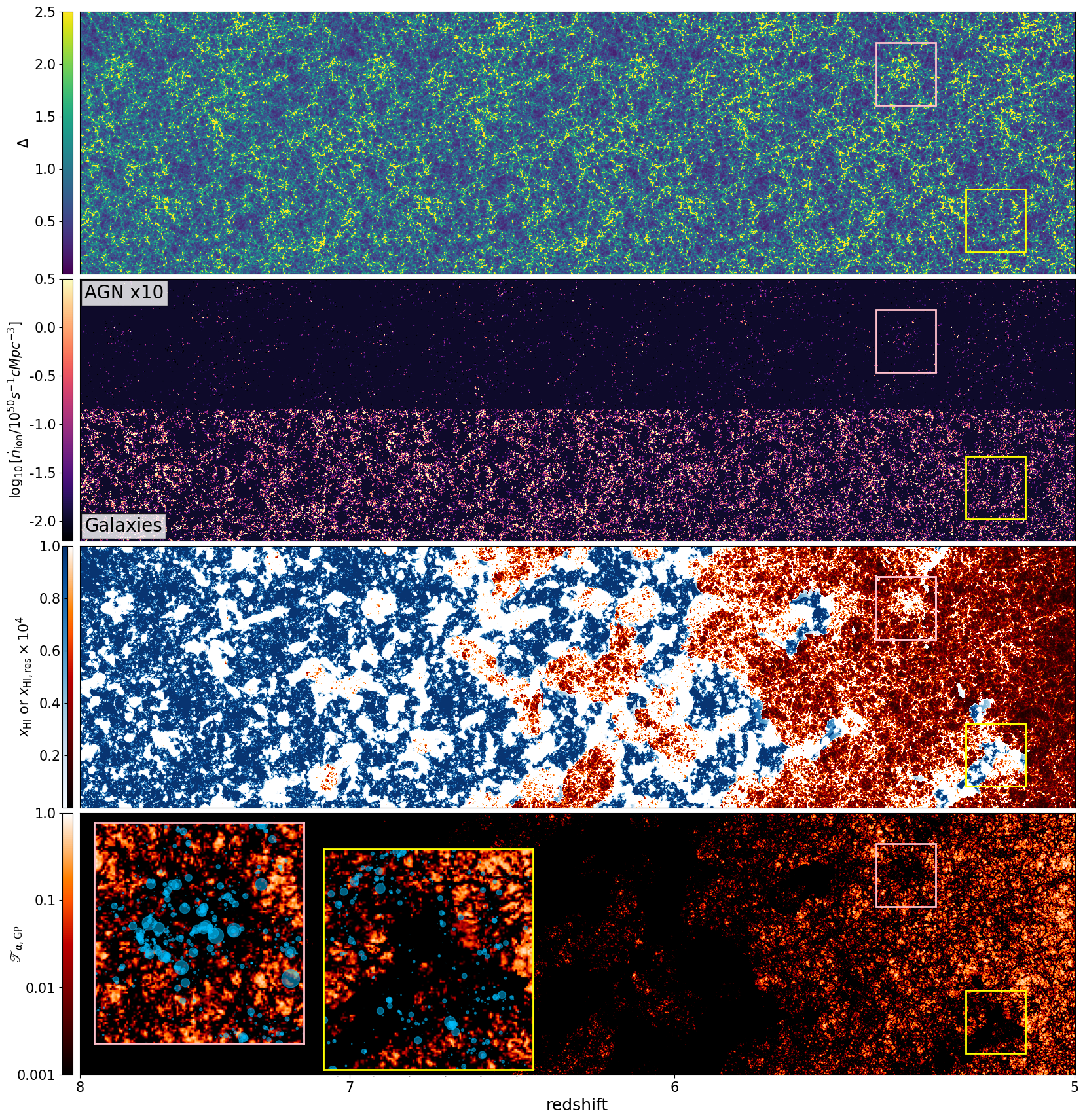} 
	\caption{Lightcones from the AGN-assisted model, with a side length of $210h^{-1}{\rm cMpc}$, a transverse depth of $410h^{-1}{\rm ckpc}$, and spanning $5\leq z\leq8$. From top to bottom, the panels show the overdensity, the ionizing emissivity from AGN and galaxies, the ionization field, and the corresponding Ly$\alpha$ transmission. In the emissivity panel, the upper half shows the AGN contribution, multiplied by a factor of 10 for better visualization, while the lower half shows the galaxy contribution. The ionization field shows neutral regions in blue and the residual neutral fraction in ionized regions is in red. The inset panels in the bottom row highlight two representative low-transmission regions, with star-forming galaxies located within a transverse distance of 10~cMpc and brighter than $M_{\rm UV}=-17$ overlaid as blue circles; larger circles correspond to brighter galaxies. These lightcones illustrate how Ly$\alpha$ transmission is shaped by the combined effects of large-scale density structure, patchy reionization, residual neutral gas in ionized regions, and the proximity of ionizing sources.}
	\label{fig:lc_qso}
\end{figure*}

We model the high-redshift galaxy population using \textsc{Meraxes}, a semi-analytic model designed to follow galaxy formation during the EoR \citep{Mutch2016,Qin2017a,Ventura2024MNRAS.529..628V}. We apply the model to dark matter halo merger trees constructed from a cosmological $N$-body simulation with a box side length of $210h^{-1}{\rm cMpc}$. The top panel of Fig.~\ref{fig:lc_qso} shows an example density lightcone from this volume, spanning $5 \leq z \leq 8$ and projected over a transverse depth of $410h^{-1}{\rm ckpc}$. It illustrates how the filamentary cosmic web becomes increasingly structured towards lower redshifts, providing the large-scale environments in which halos and ionizing sources form. Details of the halo identification and merger-tree construction are given in \citet{Balu2023MNRAS.520.3368B}. We emphasize, however, that to resolve the low-mass galaxy population expected to contribute significantly to the ionizing photon budget, we augment the merger trees\footnote{The merger-tree augmentation is necessarily approximate and can introduce mild artifacts in the galaxy catalogue. For example, the nearly vertical features visible in the 2D distributions in Fig.~\ref{fig:fesc_gal} occur around the halo mass resolution limit of the original simulation before augmentation, and likely reflect the construction of the augmented merger trees rather than physical features in the galaxy--halo relation. These artifacts are confined to a narrow range in halo mass and are not expected to affect the broad trends used in this work. We aim to improve the augmentation using generative models in future releases (\citealt{Brennan2025OJAp....8E..77B}; Brennan et al. in prep.).}
 by sampling conditional halo mass functions down to the atomic-cooling threshold at $z\leq20$ using a Monte-Carlo algorithm developed by \citet{Qiu2021MNRAS.500..493Q}.

Given these halo merger trees, \textsc{Meraxes} assigns baryonic components to dark matter halos and evolves galaxy properties through a sequence of physically motivated processes. These include gas accretion, radiative cooling, star formation, stellar evolution, supernova feedback, metal enrichment, satellite infall and disruption, galaxy mergers, blackhole growth, and heating from active galactic nuclei and photoionization. Note that \textsc{Meraxes} assumes gas to be first accreted into a hot reservoir, where it is assumed to be shock-heated to the virial temperature and distributed according to an isothermal spherical density profile. It then cools onto a rotationally supported cold-gas disc where star formation occurs. We therefore treat the cold-gas disc as an effective proxy for the star-forming interstellar medium (ISM), and the hot gas reservoir as the CGM, i.e. the halo-scale gas surrounding galaxies \citep{Tumlinson2017ARA&A..55..389T}. This interpretation does not resolve the multiphase structure of either component, but aims to provide physical motivations to connect the modelled gas reservoirs to the ionizing photon escaping models presented later.

For each galaxy, the non-ionizing UV luminosity is computed from its star-formation history using a stellar population synthesis model, with dust attenuation applied following the prescription adopted in previous \textsc{Meraxes} studies (e.g., \citealt{Qiu19}). We largely retain the galaxy-formation parameters used in \citet{Qin2023MNRAS.526.1324Q}, for which the predicted galaxy UV luminosity functions are consistent with {\it HST} and {\it JWST} measurements out to $z\sim12$. On the other hand, we compute the bolometric luminosity of an AGN assuming it always accretes at Eddington rate before the accretion disc gets fully consumed (and hence a duty cycle is introduced), and obtain its non-ionizing UV luminosity using corrections from \citet{Hopkins2007ApJ...654..731H}. More details and values for relevant parameters can be found in \citet{Qin2017a}. As the trees have changed since, we recalibrate the blackhole parameters ($k_{\rm c}=0.3$ and $k_{\rm h}=0.5$) to reproduce the observed abundance of faint AGN at high redshift. This recalibration allows us to include AGN as a physically motivated source population constrained by current observations, rather than as an unconstrained boost to the ionizing emissivity. 

\begin{figure*}
	\centering
	\includegraphics[width=.98\textwidth]{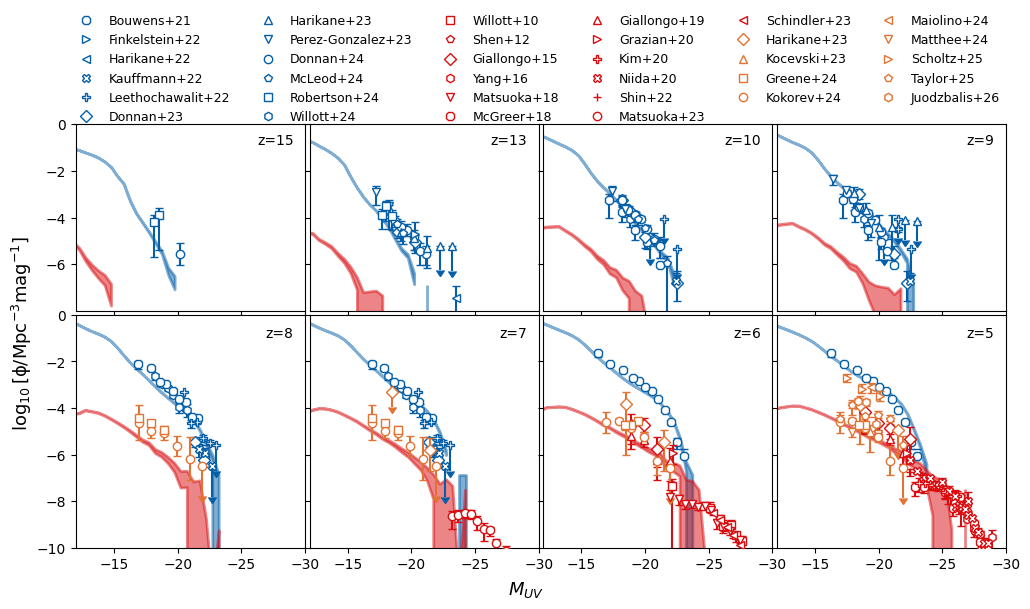}
	\caption{Galaxy (blue) and AGN (red) rest-frame UV luminosity functions predicted by our fiducial model from $z=15$ to $z=5$ with $1\sigma$ sample variance indicated. Observational measurements of galaxy \citep{Bouwens2021AJ....162...47B,Finkelstein2022ApJ...940L..55F,Harikane2022ApJ...929....1H,Kauffmann2022arXiv220711740K,Leethochawalit2022arXiv220515388L,Donnan2023MNRAS.518.6011D,Harikane2023ApJS..265....5H,PerezG2023ApJ...951L...1P,Donnan2024MNRAS.533.3222D,McLeod2024MNRAS.527.5004M,Robertson2024ApJ...970...31R,Willott2024ApJ...966...74W} and AGN UV luminosity functions \citep{Willott2010AJ....139..906W,Shen2012ApJ...746..169S,Giallongo2015,Yang2016ApJ...829...33Y,Matsuoka2018ApJ...869..150M,McGreer2018AJ....155..131M,Giallongo2019ApJ...884...19G,Grazian2020ApJ...897...94G,Kim2020ApJ...904..111K,Niida2020ApJ...904...89N} including recent JWST faint candidates \citep{Harikane2023ApJ...959...39H,Kocevski2023ApJ...954L...4K,Greene2024ApJ...964...39G,Kokorev2024ApJ...968...38K,Maiolino2024A&A...691A.145M,Matthee2024ApJ...963..129M,Scholtz2025A&A...697A.175S,Taylor2025ApJ...986..165T,Juodzbalis2026MNRAS.546ag086J} are shown for comparison. The agreement between the model and the observed source populations demonstrates that the galaxy and AGN emissivity fields used in this work are anchored to current high-redshift observations before being used to model reionization and Ly$\alpha$ forest transmission.}
	\label{fig:uvlf_agnlf}
\end{figure*}

Figure~\ref{fig:uvlf_agnlf} compares the resulting galaxy and AGN UV luminosity functions from our fiducial model (see Sec. \ref{sec:fesc}) with current observational constraints over $5 \leq z \leq 15$ (see the caption for the reference list). The calibrated model reproduces the observed galaxy UV luminosity functions across the redshift range relevant to this work, while also producing an AGN population broadly consistent with measurements of the high-redshift AGN luminosity function, including some recent faint AGN candidates identified by {\it JWST} (e.g.,  \citealt{Greene2024ApJ...964...39G,Kokorev2024ApJ...968...38K,Matthee2024ApJ...963..129M}). This illustrates that the galaxy and AGN populations entering our reionization calculations are anchored to observed high-redshift source populations. As photoheating feedback primarily affects low-mass halos below the luminosity range currently probed by observations, all models explored in this work remain consistent with the available UV luminosity-function data with minor changes in the very faint (low-mass) regime. We therefore keep the underlying galaxy- and blackhole-formation parameters fixed, and calibrate only the uncertain ionizing escape fraction ($f_{\rm esc}^{\rm gal}$) when matching the Ly$\alpha$ forest opacity distribution.

\subsection{The UV ionizing escape fraction}
Resolving the propagation of ionizing photons through the interstellar and circumgalactic environments of galaxies requires high-resolution radiation-hydrodynamic simulations (e.g.,  \citealt{Lewis2020MNRAS.496.4342L,Ma2020MNRAS.498.2001M,Rosdahl2022MNRAS.515.2386R,Kostyuk2023MNRAS.521.3077K,Yeh2023MNRAS.520.2757Y}). These simulations show that the escape fraction of ionizing photons can fluctuate strongly for individual galaxies on timescales of order Myr, while still exhibiting statistical correlations with galaxy properties. This behaviour is consistent with the expectation that the escape of ionizing photons is regulated by small-scale, bursty processes that shape the gas distribution around young stars, including feedback-driven clearing of low-column density channels through which photons can escape into the IGM.

Since such small-scale radiative-transfer processes are not resolved in our framework, we model the ionizing emissivity of each galaxy using its intrinsic stellar ionizing photon production rate and an effective escape fraction, $f_{\rm esc}^{\rm gal}$. In all models, we assume that each stellar baryon produces $N_\gamma=6000$ ionizing photons, consistent with our adopted IMF and the low-metallicity stellar populations expected at high redshift. Any uncertainty associated with this ionizing photon production efficiency is effectively absorbed in the normalization and stochasticity of $f_{\rm esc}^{\rm gal}$. To explore how different source-side prescriptions affect the inferred reionization history, we consider several parameterizations in which $f_{\rm esc}^{\rm gal}$ scales with host halo mass ($M_{\rm vir}$), stellar mass ($M_*$), star formation rate (SFR), specific star formation rate (sSFR), and star-forming disc surface density ($\Sigma_{\rm SF}$).

In the AGN-assisted model, we additionally include ionizing photons produced by accreting blackholes following the approach of \citet{Qin2017a} with minor modifications. For each accretion episode, we compute the intrinsic AGN ionizing emissivity from the non-ionizing UV luminosity evaluated at the midpoint of the accretion event. We assume a broken power-law spectral energy distribution, with spectral indices of $0.44$ and $1.57$ at wavelengths longer and shorter than 1200~\AA, respectively, and integrate the spectrum blueward of the Lyman limit to estimate the ionizing photon production rate. Because the AGN luminosity evolves non-linearly during an accretion episode, partly due to the bolometric correction, we use this midpoint emissivity as a representative value during active accretion. The total number of ionizing photons contributed by the accretion episode is then estimated by multiplying this emissivity by the accretion duration. However, when computing the photoionization rate and the residual neutral fraction, the relevant quantity is the effective instantaneous emissivity averaged over the simulation snapshot. We therefore weight the midpoint emissivity by the accretion duty cycle, since the AGN emissivity is only active during the fraction of the snapshot in which the blackhole is accreting\footnote{To validate this approximation, we also tested a stochastic accretion model in which each blackhole begins accreting at a random time within a snapshot and continues into the next snapshot if material remains in the accretion disc. If the disc is exhausted before the end of a snapshot, the ionizing emissivity is allowed to decay exponentially on a relaxation timescale set by the inverse of the sum of the \textit{local} photoionization and recombination rates (i.e., $t_{\rm eq}^{-1}\sim \Gamma_{\rm ion}+\Gamma_{\rm rec})$. We find that this relaxation is short compared with the snapshot spacing, so the stochastic treatment effectively reduces to an on--off accretion model. Averaged over the blackhole population, it gives ionizing emissivities consistent with our default duty-cycle-weighted prescription.}. We assume that ionizing photons produced by AGN escape\footnote{\label{footnote:fescAGN}Following \citet{Qin2017a}, we include an obscuration factor of ${\sim}23.4$\% when computing the observable AGN UV luminosity function, and apply the same factor to the AGN ionizing emissivity as an angle-averaged approximation. A more stochastic treatment of obscuration would be assigning each AGN randomly as either unobscured or obscured, contributing ionizing photons only in the former case. This would leave the AGN luminosity function and mean emissivity unchanged, but would increase source stochasticity and could enhance spatial fluctuations in the ionizing background, especially because AGN are rare and biased tracers. We do not model this additional stochasticity here as its impact is limited for our calibrated model, where AGN remain subdominant. Similar assumptions linking AGN ionizing escape to the unobscured fraction are adopted in AGN reionization models \citep[e.g.][]{Trebitsch2023MNRAS.518.3576T}, while direct measurements at $z\sim3$--4 also suggest average AGN LyC escape fractions below unity \citep{Iwata2022MNRAS.509.1820I}.} into the IGM with $f_{\rm esc}^{\rm AGN}=1$. This assumption applies only to photons emitted by the AGN component. Ionizing photons produced by stars in the same host galaxy are still attenuated by the stellar escape fraction described earlier. This distinction reflects the fact that AGN radiation originates from the nuclear region and may escape through different, potentially more highly ionized channels than stellar ionizing photons produced across the star-forming ISM.

In the CGM-attenuated models, we modify the galaxy escape fraction by applying an additional source-side attenuation on circumgalactic scales. This effect is distinct from the photoheating feedback \citep{Sobacchi2013MNRAS.432L..51S} already included in \textsc{Meraxes}, which reduces the ability of low-mass halos to accrete gas and form stars after they are exposed to the ionizing background, thereby regulating the production of ionizing photons. In contrast, the CGM attenuation considered here acts on photons after they have been produced and have escaped the star-forming ISM, but before they reach the wider IGM. We assume that these photons may be further absorbed by dense gas structures in the CGM, with the strength of this attenuation rapidly increasing with the local clumping factor ($C$). The physical motivation is that, toward the late stages of reionization, halo environments may contain increasingly prominent small-scale gas structure as galaxies grow, accrete gas, and experience feedback. Compared to a smooth CGM, these dense structures have higher recombination rates and larger neutral columns densities, making them more effective at absorbing ionizing photons \citep{Reddy2016ApJ...828..108R,Faucher-G2016MNRAS.461L..32F,Stern2021MNRAS.507.2869S}. The CGM-attenuated models therefore phenomenologically capture the possibility that late-time growth in circumgalactic clumpiness increases the opacity around galaxies and reduces the fraction of stellar ionizing photons that reach the IGM.

\subsection{The ionization fields}\label{subsec:ionization}

The escaping ionizing emissivity\footnote{When computing the ionization fields, including the Ly$\alpha$ forest optical depths, we use the instantaneous star formation rate to determine both the ionizing emissivity, $\dot{n}_{\rm ion}$, and the local photoionization rate, $\Gamma_{\rm ion}$, and adopt the post-EoR mean-free-path treatment of \citetalias{Qin2025PASA...42...49Q}. However, when estimating photoheating feedback on galaxy formation, we follow \citet{Mutch2016} and use the star formation rate averaged over the Hubble time. This is because the filtering-mass prescription \citep{Sobacchi2013MNRAS.432L..51S} assumes a slowly varying ionizing background within each cell, and is not intended to respond to short-timescale, bursty fluctuations in star formation. The Hubble-time-averaged star formation rate therefore provides an effective UVB estimate for suppressing gas accretion in low-mass halos.} from the \textsc{Meraxes} galaxy and AGN catalogue is gridded onto the cosmological volume and used to evolve the ionization state of the IGM with a customized version of \textsc{21cmFAST} \citep{Mesinger2007ApJ...669..663M,Mesinger2011MNRAS.411..955M,Murray2020}. In this excursion-set framework, a region is flagged as ionized if the cumulative number of ionizing photons produced by galaxies and AGN (after applying the relevant escape fractions) exceeds the number of hydrogen atoms after accounting for recombinations \citep{Sobacchi2014MNRAS.440.1662S}. 
We illustrate this procedure for the AGN-assisted model (see more in Sec.~\ref{sec:fesc}) in Fig.~\ref{fig:lc_qso}. The emissivity lightcones show that both galaxies and AGN trace the underlying large-scale structure, with the galaxy contribution more broadly distributed and the AGN contribution concentrated around rarer, highly biased systems. The resulting ionization field follows these source distributions: ionized regions grow first around clustered emissivity peaks and then expand into the surrounding IGM, leaving large underdense regions neutral until later times. This produces the expected inside-out morphology of reionization, with ionized bubbles growing around clustered sources before gradually percolating through the volume \citep[e.g.][]{Iliev2008tera.conf...31I,Trac2011ASL.....4..228T}.

Following \citetalias{Qin2025PASA...42...49Q} and \citet{Sobacchi2014MNRAS.440.1662S}, we also account for unresolved density structure by adopting the sub-grid density distribution of \citet{Miralda2000ApJ...530....1M}. Within each resolved cell (of ${\sim}410\ h^{-1}$~ckpc in this work), the gas is described by a volume-weighted probability distribution, $P_{\rm V}(\Delta_{\rm sub})$, where $\Delta_{\rm sub}$ is the sub-grid overdensity relative to the cell-averaged hydrogen density ($n_{\rm H}$). Note that this distribution varies with redshift and $n_{\rm H}$. We then integrate over $P_{\rm V}$ to compute the average residual neutral fraction ($x_{\rm HI, res}$), recombination rate ($\Gamma_{\rm rec}$), and clumping factor ($C$) of each cell:
\begin{equation}\label{eq:xhi_res}
	x_{\rm HI, res} = \int {\rm d}\Delta_{\rm sub} P_{\rm V} x_{\rm HI, sub};
\end{equation}
\begin{equation}
	\Gamma_{\rm rec} = n_{\rm H}	\int {\rm d}\Delta_{\rm sub} P_{\rm V} \Delta_{\rm sub}  \alpha_{\rm B}\left(1-x_{\rm HI,sub}\right)^2;
\end{equation}
and
\begin{equation}
	C = \frac{\int {\rm d}\Delta_{\rm sub} P_{\rm V} \Delta_{\rm sub}^2\left(1-x_{\rm HI, sub}\right)^2 }{\left[\int {\rm d}\Delta_{\rm sub} P_{\rm V} \Delta_{\rm sub} \left(1-x_{\rm HI,sub}\right)\right]^2}.
\end{equation}
Here, $x_{\rm HI,sub}$ is the neutral fraction at sub-grid density and computed assuming the ionized region is at photoionization equilibrium,
\begin{equation}
	x_{\rm HI,sub} f_{\rm ion,ss} \Gamma_{\rm ion} =
	\chi_{\rm HeII} n_{\rm H}\Delta_{\rm sub} (1-x_{\rm HI,sub})^2 \alpha_{\rm B},
	\label{eq:equilibirum}
\end{equation}
where $\chi_{\rm HeII}$ accounts for singly ionized helium, $\Gamma_{\rm ion}$ represents the local photoionization rate, $f_{\rm ion,ss}\left(\Delta_{\rm sub}, \Gamma_{\rm ion}, T, z\right)$ accounts for self-shielding attenuation \citep{Schaye2001ApJ...559..507S,Rahmati2013MNRAS.430.2427R} and $\alpha_{\rm B}(T)$ is the recombination coefficient at gas temperature $T$. This sub-grid treatment is important as dense gas contributes disproportionately to recombinations and self-shielded absorption even when it occupies only a small fraction of the cell volume. Fig.~\ref{fig:lc_qso} also illustrates the resulting residual neutral fraction field in ionized regions: although large connected regions are flagged as ionized by the excursion-set calculation, the residual neutral fraction remains spatially inhomogeneous and is enhanced in dense structures.

The ionized gas temperature\footnote{We ignore pre-heating by X-rays in this work and set the neutral gas temperature using fluctuating adiabatic cooling \citep{Seager1999ApJ...523L...1S,Munoz2023MNRAS.523.2587M}. For partially ionized cells, we assign a volume-weighted temperature from their ionized and neutral components. However, both neutral and partially ionized cells are treated as completely opaque to Ly$\alpha$ photons, so finite transmission arises only from ionized regions, where the residual neutral fraction is set by photoionization equilibrium.} is evolved using the post-reionization thermal model of \citet[][see also \citetalias{Qin2025PASA...42...49Q}]{McQuinn2016MNRAS.456...47M}, in which each cell retains memory of its reionization redshift. When a cell is first reionized, the gas gets heated impulsively by supersonic ionization fronts to $2\times10^4~{\rm K}$ \citep{DAloisio2019ApJ...874..154D}. It then cools gradually toward a relaxation temperature, which we set to $250(1+z)$~K at the mean density and scale with density as $n_{\rm H}^{1/\gamma}$, where the equation-of-state index is fixed to $\gamma=1.7$. Regions ionized more recently are therefore generally hotter, while regions ionized earlier have had more time to cool through adiabatic expansion and Compton cooling \citep{McQuinn2016MNRAS.456...47M}.

\subsection{Ly$\alpha$ forest optical depths}

\begin{table*}
	\centering
	\caption{Models explored in this work. The fiducial and following two models use an escape fraction parameterized by halo mass for galaxies, while the remaining models scale $f_{\rm esc}^{\rm gal}$ against different galaxy properties including stellar mass, star formation rate and their ratio, as well as the surface density of star-forming disc. The AGN-assisted model includes an additional AGN contribution to the ionizing emissivity, whereas the CGM-attenuated and SF Disc models include additional suppression of $f_{\rm esc}^{\rm gal}$ from CGM attenuation. The resulting CMB optical depths and colours used to identify these models in figures are also listed.}
	\label{tab:model_suite}
	\renewcommand{\arraystretch}{1.5}
	
	\begin{tabular}{l|l|c|p{1mm}l}
		\hline
		\hline
		Model & $f_{\rm esc}^{\rm gal}$ parameterization & $\tau_e$ & \multicolumn{2}{l}{Description}  \\	
		\hline
		Fiducial
		& $0.08\left(M_{\rm vir}/10^{10}{\rm M}_\odot\right)^{-0.3}$ &0.0636
		& \modelcolorbox{fiducialcol}& \ $f_{\rm esc}^{\rm gal}$ depends on host halo mass and increases towards lower masses\\
		\hline
		AGN-assisted 
		&$0.06\left(M_{\rm vir}/10^{10}{\rm M}_\odot\right)^{-0.4}$ (1 for AGN) &0.0652
		& \modelcolorbox{agncol}& \ Similar to Fiducial but w/ more contribution from lower-mass halos \& AGN \\
		\hline
		CGM-attenuated
		&  
		\begin{tabular}[c]{@{}l@{}}$0.1\left(M_{\rm vir}/10^{10}{\rm M}_\odot\right)^{-0.2} \times $\\$ \exp\left[-\left(\Sigma_{\rm CGM} / \,{\rm M}_{\odot}\,{\rm pc}^{-2}\right)^{0.2} (C/6.3)^3\right]$
		\end{tabular} & 0.0609
		& 	\modelcolorbox{cgmcol}&\renewcommand{\arraystretch}{1.0}
		\begin{tabular}[c]{@{}l@{}}Similar to Fiducial but w/ more contribution from higher-mass halos \&\\[-0.2ex] $f_{\rm esc}^{\rm gal}$ is further suppressed in CGM whose opacity increases w/ column\\[-0.2ex] density and clumping factor
		\end{tabular} \\
		\hline
		Stellar Mass
		& $0.073\left(M_*/10^{8}{\rm M}_{\odot}\right)^{-0.2}$ & 0.0596
		& \modelcolorbox{mstarcol}& \ $f_{\rm esc}^{\rm gal}$ depends on stellar mass\\
		\hline
		SFR
		& $0.06\left({\rm SFR}/1\,{\rm M}_{\odot}\,{\rm yr}^{-1}\right)^{-0.33}$ &0.0571
		& \modelcolorbox{sfrcol} & \ $f_{\rm esc}^{\rm gal}$ depends on star formation rate\\
		\hline
		sSFR
		& $0.055\left({\rm sSFR}/10\,{\rm Gyr}^{-1}\right)^{0.65}$& 0.0652 
		& \modelcolorbox{ssfrcol} & \ $f_{\rm esc}^{\rm gal}$ depends on specific star formation rate\\
		\hline
		SF Disc
		&
		\begin{tabular}[c]{@{}l@{}}$0.063\left(\Sigma_{\rm SF}/1{\rm M}_{\odot}\,{\rm pc}^{-2}\right)^{0.12}\times$\\$ \exp\left[-\left(\Sigma_{\rm CGM} / \,{\rm M}_{\odot}\,{\rm pc}^{-2}\right)^{0.2} (C/6.3)^3\right]$
		\end{tabular}
		&0.0535& \modelcolorbox{sfdisccol}& \ $f_{\rm esc}^{\rm gal}$ depends on star-forming disc surface density w/ CGM suppression \\
		\hline
		
	\end{tabular}
\end{table*}

We compute the Ly$\alpha$ forest transmission from the simulated IGM fields using a forward-modelling procedure similar\footnote{Unlike \citetalias{Qin2025PASA...42...49Q} where the residual neutral fraction was computed using equation~(\ref{eq:equilibirum}) with cell-averaged properties, this work uses $x_{\rm HI,res}$ evaluated from equation~(\ref{eq:xhi_res}) based on the sub-grid density model described in Sec \ref{subsec:ionization}. This also provides a more consistent input optical depth for the stochastic mapping mentioned next.} to that of \citetalias{Qin2025PASA...42...49Q}. For each cell in ionized regions, we first estimate the Ly$\alpha$ optical depth using the fluctuating Gunn--Peterson approximation (FGPA; \citealt{gunn1965density,Weinberg1999elss.conf..346W}):
\begin{equation}
	\tau_{\alpha,{\rm GP}}
	=
	\sqrt{\frac{3\pi\sigma_{\rm T}}{8}}
	c f_{\alpha}\lambda_{\alpha}
	H^{-1}(z)
	n_{\rm H}x_{\rm HI,res},
	\label{eq:tau_gp}
\end{equation}
where $\sigma_{\rm T}$ is the Thomson cross-section, $c$ is the speed of light, $f_{\alpha}=0.416$ is the Ly$\alpha$ oscillator strength, $\lambda_{\alpha}=1216$ {\AA} is the Ly$\alpha$ rest-frame wavelength, and $H(z)$ is the Hubble parameter. Fig.~\ref{fig:lc_qso} shows the corresponding Ly$\alpha$ transmission lightcone from the our model: neutral regions are completely opaque, while transmission through ionized regions varies strongly with the density field and the residual neutral fraction. The resulting transmission is therefore highly intermittent, with transparent segments appearing only along sightlines that pass through sufficiently ionized, low-opacity regions. On the other hand, low Ly$\alpha$ transmission can arise from different environments as highlighted by the two inset panels. One example lies in an overdense region containing a clustered galaxy population, where the opacity is enhanced by dense residual neutral gas despite the proximity of ionizing sources. The other lies in a more underdense region with few nearby galaxies, where weak local emissivity and delayed ionization can also produce a low-transmission segment \citep{Kakiichi2018MNRAS.479...43K,Kashino2026ApJ...997..280K,Zhu2026ApJ..1002...93Z}. Thus, opaque Ly$\alpha$ troughs do not map uniquely onto either overdensities or voids, but instead reflect the combined effects of density, source clustering, ionization topology, and residual neutral gas.

The FGPA provides a cell-level estimate of the Ly$\alpha$ opacity, but it does not explicitly model the full Ly$\alpha$ line profile, thermal broadening, peculiar velocities, or the unresolved small-scale gas structure that contributes to the observed Ly$\alpha$ forest. We therefore follow \citetalias{Qin2025PASA...42...49Q} and correct the raw FGPA optical depths using a stochastic mapping calibrated against higher-resolution hydrodynamical simulations \citep{Bolton2017MNRAS.464..897B}. We refer interested readers to \citetalias{Qin2025PASA...42...49Q} for the details of this procedure, and note only that we recompute the mapping for the spatial scale relevant to this work.

With the corrected optical depth ($\tau_{\alpha}$), we then compute the transmitted flux ($F\equiv e^{-\tau_{\alpha}}$). To compare with observations, we interpolate the simulation snapshots in comoving distance to construct Ly$\alpha$ transmission lightcones, from which we draw random sightlines to generate mock spectra. We then compress these spectra into the effective optical depth:
\begin{equation}
	\tau_{\rm eff}
	=
	-\ln \langle F\rangle,
	\label{eq:tau_eff}
\end{equation}
where $\langle F\rangle$ is the mean transmitted flux measured over the same redshift range (or length) as the observational data ($\Delta z=0.1$; \citealt{Bosman2022MNRAS.514...55B}).

\section{From escape fractions to ionizing source populations}\label{sec:fesc}

Table~\ref{tab:model_suite} summarizes the model variants explored in this work. In all cases, we keep the underlying \textsc{Meraxes} galaxy- and blackhole-formation parameters fixed, and vary only the prescription for the galaxy escape fraction, $f_{\rm esc}^{\rm gal}$ (except in the AGN-assisted model where accreting blackholes also contribute ionizing photons). We parameterize $f_{\rm esc}^{\rm gal}$ as a power law of either halo mass or a galaxy property, $f_{\rm esc}^{\rm gal}=\min\left[1, f_0 \left({X}/{X_0}\right)^{\alpha}\right]$, where $X$ is the chosen quantity, $X_0$ is a fixed pivot value, and $f_0$ and $\alpha$ are the normalization and scaling index. The pivot values are chosen to lie in the range most relevant for the galaxy population contributing to reionization, guided by the relation between halo mass and galaxy properties shown in the upper panels of Fig.~\ref{fig:fesc_gal}. For each choice of $X$, we perform a grid search over $f_0$ and $\alpha$, generate the corresponding ionization fields and Ly$\alpha$ forest optical depths using the method described above, and select the model that best matches\footnote{Because each model requires running the full \textsc{Meraxes} galaxy catalogue and reionization calculation, the grid search is computationally expensive. A single model evaluation takes ${\sim}150$ CPU hours using 96 Intel(R) Xeon(R) Platinum 8470Q cores and requires ${\sim}280{\rm GB}$ of memory. The best-fitting parameter values quoted here should therefore not be interpreted as fully optimized global optima. However, the grid is sufficient for our purpose: models with the opposite sign of the scaling index give substantially poorer fits to the Ly$\alpha$ forest data, so the preferred qualitative trends are robust.} the observed cumulative distribution of Ly$\alpha$ effective optical depths from \citet{Bosman2022MNRAS.514...55B}.

The middle and bottom panels of Fig.~\ref{fig:fesc_gal} show the resulting $f_{\rm esc}^{\rm gal}$ relation from these prescriptions and how they change the halo masses that dominate the ionizing photon budget. For each model, we plot the cumulative fraction of ionizing photons produced by halos above a given $M_{\rm vir}$, i.e. $f_{\rm ion}(>M_{\rm vir})$, both before and after applying $f_{\rm esc}^{\rm gal}$. Since the underlying \textsc{Meraxes} galaxy population is fixed across the model suite, the intrinsic cumulative distributions are nearly identical at fixed redshift. Their redshift evolution is instead driven by galaxy formation itself: towards lower redshift, more massive halos host an increasing fraction of the star formation, so the intrinsic ionizing photon budget shifts progressively towards higher halo masses. Differences between models therefore arise mainly from how the escape-fraction prescription redistributes this intrinsic photon budget.

\subsection{The fiducial model}

\begin{figure*}
	\centering
	\includegraphics[width=\textwidth]{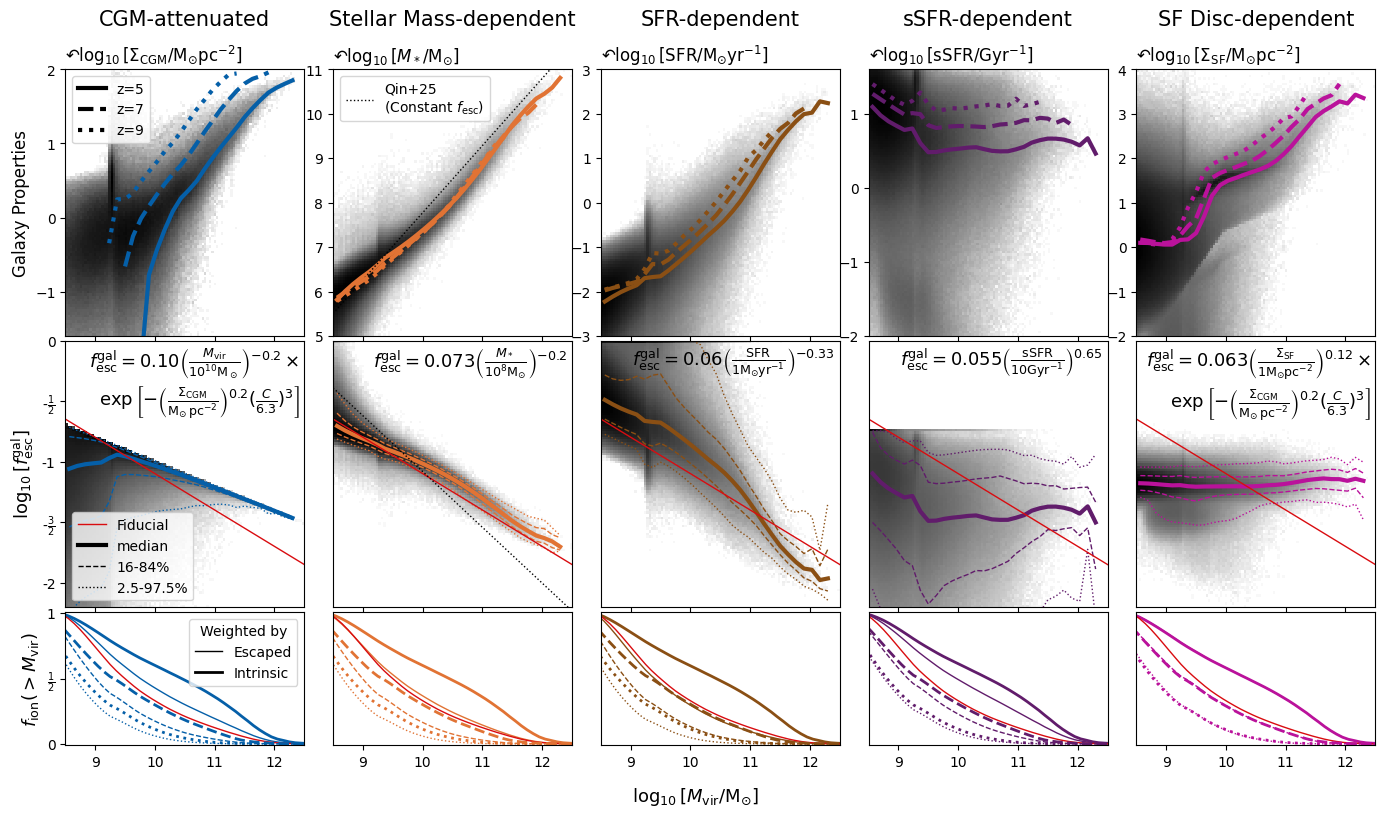}
	\caption{Relations between host halo mass, galaxy properties, escape fractions, and ionizing photon contributions for the model variants explored in this work.
		{\it Top panels}: properties of star-forming galaxies used in the escape-fraction prescriptions, including  CGM column density, stellar mass, star-formation rate, specific star-formation rate, and star-forming disc surface density. Solid, dashed, and dotted curves show the median relations at $z=5$, 7 and 9, respectively, spanning the redshift range over which most of reionization occurs. The shaded regions show the two-dimensional distribution of simulated galaxies at $z=5$, and stellar (and escape fraction) to halo mass ratio of the best-fitting of Constant-$f_{\rm esc}$ model from \citetalias{Qin2025PASA...42...49Q} is shown as black dotted lines for comparison.
		{\it Middle panels}: corresponding galaxy escape fractions, $f_{\rm esc}^{\rm gal}$, as a function of $M_{\rm vir}$. Thick curves show the median relation at $z=5$, while the inner and outer thin curves enclose the [16, 84]th and [2.5, 97.5]th percentiles, respectively. The fiducial halo-mass-dependent escape-fraction model is shown as red solid lines for comparison.
		{\it Bottom panels}: cumulative fraction of ionizing photons produced by halos above a given $M_{\rm vir}$. Different line styles correspond to the same redshifts as in the top panels, while line thickness distinguishes the intrinsic ionizing photon production (thick) from the escaping contribution after applying $f_{\rm esc}^{\rm gal}$ (thin). 
		We show the fiducial model at $z=5$ and weighted by escaped budget for comparison.
		This figure illustrates how tying $f_{\rm esc}^{\rm gal}$ to galaxy properties, rather than directly to halo mass, introduces scatter in the ionizing emissivity at fixed $M_{\rm vir}$ and changes which halo masses dominate the escaping ionizing photon budget.}
	\label{fig:fesc_gal}
\end{figure*}

The fiducial model takes $(X_0\equiv M_{\rm vir}, f_0, \alpha)=(10^{10}{\rm M}_{\odot}, 0.08, -0.3)$, so that $f_{\rm esc}^{\rm gal}$ increases towards lower halo masses. This is broadly consistent with the conclusion of \citetalias{Qin2025PASA...42...49Q}, where the Ly$\alpha$ forest data favoured reionization histories in which the ionizing emissivity is dominated by relatively low-mass halos, with $\alpha=-0.46_{-0.13}^{+0.09}$ for their Constant-$f_{\rm esc}$ model\footnote{The Constant-$f_{\rm esc}$ model in \citetalias{Qin2025PASA...42...49Q} uses the same halo-mass-dependent definition of $f_{\rm esc}^{\rm gal}$ as our fiducial model. However, the two models differ in their treatment of the stellar-to-halo mass relation and the star-formation duty cycle.}. In this picture, low-mass galaxies help produce a more gradual and extended EoR, rather than a rapid transition driven mainly by rare massive sources. This is also visible in the bottom panels of Fig.~\ref{fig:fesc_gal}: applying the fiducial escape fraction shifts the escaping ionizing photon budget towards lower halo masses relative to the intrinsic photon production. We therefore use this model as the baseline case in this work. The remaining model variants test whether, after calibration to the same Ly$\alpha$ forest data, the allowed EoR history broadens when the ionizing emissivity is modified by AGN contribution, CGM attenuation, or evolving galaxy properties rather than being tied to halo mass alone.

\subsection{The AGN-assisted model}
The AGN-assisted model also uses a halo-mass-dependent stellar escape fraction, but adds ionizing photons from accreting blackholes. Since the AGN emissivity is set by the \textsc{Meraxes} blackhole population, which is calibrated to reproduce the observed AGN UV luminosity functions, the grid search adjusts only the stellar component around this additional contribution. The best-fitting stellar prescription has $(f_0,\alpha)=(0.06,-0.4)$, i.e. a steeper halo-mass dependence than the fiducial model. This shifts the stellar ionizing emissivity further towards lower-mass halos, while the AGN component provides an additional contribution that becomes more important towards lower redshifts as the blackhole population grows.

\subsection{The CGM-attenuated model}
The CGM-attenuated model is designed to explore the opposite direction. Instead of allowing the emissivity to become even more weighted towards low-mass halos, we ask whether a model with a relatively larger contribution from higher-mass halos can also match the Ly$\alpha$ forest once additional source-side attenuation is included. We therefore adopt a shallower intrinsic halo-mass dependence, fixing $\alpha=-0.2$, compared with $\alpha=-0.3$ in the fiducial model and $\alpha=-0.4$ in the AGN-assisted model. We then include an additional CGM attenuation term based on the CGM column-density proxy, $\Sigma_{\rm CGM}$, and the local clumping factor, $C$. 

As shown in the first column of Fig.~\ref{fig:fesc_gal}, $\Sigma_{\rm CGM}$ increases on average with halo mass, reflecting the larger hot-gas reservoirs of more massive halos, but with appreciable scatter due to variations in gas content and assembly history. The weak mass dependence of $\Sigma_{\rm CGM}$ mainly modulates the intrinsic halo-mass trend, while the steep dependence on $C$ makes the attenuation most important in dense environments and at late times. The resulting $f_{\rm esc}^{\rm gal}$ is therefore much less enhanced towards low masses, and in some cases even suppressed, with a scatter of order $\sim0.3$--$0.5$ dex in the 16th--84th percentile range of $\log_{10} f_{\rm esc}^{\rm gal}$ at fixed halo mass. Correspondingly, the bottom leftmost panel shows a weaker shift towards low-mass halos than in the fiducial model: the escaping photon budget remains closer to the intrinsic, more massive-halo-weighted distribution. In practice, we find that a steep clumping dependence, taken here to be $C^3$, is needed to reduce the escaping stellar emissivity towards the end of reionization. We discuss the impact of this choice on the emissivity and reionization history in Sec.~\ref{sec:robust_eor}.

\subsection{The stellar-mass-dependent model}
The remaining models tie $f_{\rm esc}^{\rm gal}$ to galaxy properties predicted by \textsc{Meraxes}. Fig.~\ref{fig:fesc_gal} also shows how these quantities map onto halo mass, ordered from left to right by the CGM column-density proxy discussed above, stellar mass, SFR, sSFR, and star-forming disc surface density. The stellar-mass-dependent model is the closest galaxy-property analogue of a smooth halo-mass-dependent prescription, because $M_*$ is strongly correlated with $M_{\rm vir}$ over the mass range relevant here. With $(X_0 \equiv M_*, f_0, \alpha)=(10^8{\rm M}_\odot, 0.073, -0.2)$, the median $f_{\rm esc}^{\rm gal}$--$M_{\rm vir}$ relation is very similar to the fiducial model. The main difference is the small amount of scatter inherited from the stellar-to-halo-mass relation: at fixed halo mass, the 16th--84th percentile range in $\log_{10} f_{\rm esc}^{\rm gal}$ is typically $\sim0.1$--$0.2$ dex, while the 2.5th--97.5th percentile range is up to $0.4$ dex. Because the effective escape-fraction relation is so similar to the fiducial case, the escape fraction-weighted cumulative photon budget also closely follows the fiducial low-mass shift in the bottom row.

\subsection{The SFR-dependent model}
The SFR-dependent model also behaves similarly to the fiducial one. Since SFR increases strongly with $M_{\rm vir}$, the negative scaling selected by the grid search, $(X_0\equiv {\rm SFR}, f_0,\alpha)=(1{\rm M}_\odot {\rm yr}^{-1}, 0.06, -0.33)$, maps into an escape fraction that also decreases with halo mass. The resulting median $f_{\rm esc}^{\rm gal}$--$M_{\rm vir}$ relation therefore closely follows the fiducial model over most of the relevant halo-mass range, but with more scatter than the stellar-mass-dependent case. At fixed halo mass, the 16th--84th percentile range in $\log_{10} f_{\rm esc}^{\rm gal}$ is typically of order half a dex over much of the relevant mass range. Unlike the fiducial and stellar-mass-dependent models, however, the effective halo-mass weighting evolves more noticeably with redshift. At fixed $M_{\rm vir}$, galaxies have larger SFRs at higher redshifts, so the negative SFR scaling assigns lower escape fractions to the same halo mass at earlier times. This strengthens the relative contrast between low- and high-mass systems and produces a larger shift of the escaping photon budget towards lower halo masses in the earlier universe. The lower middle panel of Fig.~\ref{fig:fesc_gal} therefore shows a stronger redshift dependence in the SFR-dependent redistribution than in the fiducial or stellar-mass-dependent models, even though the $z=5$ median relation remains close to the fiducial case.

\subsection{The sSFR-dependent model}
The sSFR-dependent model has the weakest monotonic relation with halo mass among these galaxy properties. The upper panel shows that sSFR depends only weakly on $M_{\rm vir}$ and has substantial scatter, reflecting the bursty nature of star formation in high-redshift galaxies. For this model, the grid search selects $(X_0 \equiv {\rm sSFR},f_0,\alpha)=(10{\rm Gyr}^{-1}, 0.055,0.65)$, i.e. a positive scaling of $f_{\rm esc}^{\rm gal}$ with sSFR. At low halo masses, $M_{\rm vir}\lesssim5\times10^9{\rm M}_{\odot}$, this still maps into an effective $f_{\rm esc}^{\rm gal}$--$M_{\rm vir}$ relation similar to the fiducial model but with a much lower contribution. At higher masses, however, the relation becomes much flatter because sSFR itself is nearly flat with halo mass. The resulting escape fractions are therefore much more broadly distributed than in a direct halo-mass-dependent prescription, with the 16th--84th percentile range in $\log_{10} f_{\rm esc}^{\rm gal}$ spanning of order $\sim1$ dex over much of the relevant mass range. Its $f_{\rm ion}(>M_{\rm vir})$ panel in Fig.~\ref{fig:fesc_gal} shows that applying the sSFR-dependent escape fraction does not strongly redistribute the ionizing photon budget towards lower-mass halos, and the escape-weighted cumulative distribution remains closer to the intrinsic distribution than in the fiducial model. We note that the upper envelope visible in its $f_{\rm esc}^{\rm gal}$ panel reflects the finite simulation timestep, which limits the maximum sSFR and therefore caps the corresponding $f_{\rm esc}^{\rm gal}$.

\subsection{The star-forming-disc model}
The star-forming-disc model takes $X=\Sigma_{\rm SF}$. This quantity generally increases with halo mass, but with significant scatter, especially towards lower masses. The upper panel also shows that $\Sigma_{\rm SF}$ becomes relatively flat at $M_{\rm vir}\lesssim10^{10}{\rm M}_{\odot}$, with a drop around $M_{\rm vir}\sim5\times10^9$--$10^{10}{\rm M}{\odot}$ where galaxies are only marginally able to sustain sufficient gas for star formation against the boosted supernova feedback \citep{Qiu19}. Because of this flat low-mass relation, changing the power-law parameters with $\Sigma_{\rm SF}$ cannot strongly differentiate between halos below $\sim10^{10}{\rm M}_{\odot}$: it can change the normalization of the escape fraction for the low-mass population as a whole, but it does not produce a strong relative weighting towards the lowest-mass halos. We therefore find that a pure $\Sigma_{\rm SF}$-dependent escape-fraction model does not provide a satisfactory match to the Ly$\alpha$ forest CDFs.

The model listed in Table~\ref{tab:model_suite} instead adopts the same CGM attenuation prescription as the halo-mass-dependent CGM-attenuated model, and should be interpreted as an SF-disc-dependent intrinsic escape fraction with additional source-side attenuation. The grid search selects a weak positive scaling, with $(X_0=\Sigma_{\rm SF}, f_0,\alpha)=(1{\rm M}_\odot {\rm pc}^{-2}, 0.063,0.12)$, so that galaxies with higher star-forming surface densities are assigned slightly larger intrinsic escape fractions\footnote{The weak positive scaling with $\Sigma_{\rm SF}$ should be interpreted cautiously. Although larger gas columns would naively be expected to reduce the escape of ionizing photons, $\Sigma_{\rm SF}$ is a global star-forming disc quantity in our model, rather than the local neutral column density along the photon escaping paths. It may therefore also trace clustered star formation and feedback-driven porosity in the ISM. We also identified a candidate model with a negative scaling, $(f_0,\alpha)=(0.55,-0.40)$, which is more directly aligned with the absorption expectation, but this model gives a poorer match to the Ly$\alpha$ forest CDFs than the selected SF-disc model.} before the additional CGM attenuation is applied. After applying CGM attenuation, the resulting escape fractions retain modest scatter at fixed $M_{\rm vir}$: the 16th--84th percentile range in $\log_{10} f_{\rm esc}^{\rm gal}$ is typically ${\sim}0.2$ dex, while the 2.5th--97.5th percentile range can reach ${\sim}0.5$ dex. The resulting $f_{\rm esc}^{\rm gal}$--$M_{\rm vir}$ relation is also close to flat over the relevant mass range and, as the escape fraction does not strongly reweight the photon budget across halo mass, the intrinsic and escape-weighted cumulative photon distributions in the bottom rightmost panel highly overlap with each other. The SF-disc model therefore remains the most massive-halo-weighted model in our suite. 

\subsection{The UV luminosities of ionizing sources}
\begin{figure}
	\centering
	\includegraphics[width=\columnwidth]{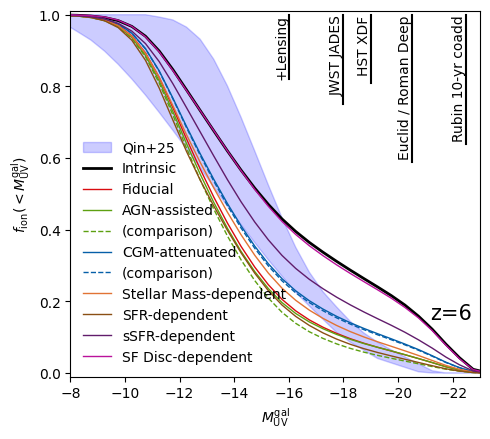}
	\caption{Cumulative fraction of ionizing photons contributed by galaxies brighter than a given UV magnitude at $z=6$. The black curve shows the intrinsic ionizing photon production from the \textsc{Meraxes} galaxy population before applying escape-fraction weighting. Coloured curves show the escaping ionizing photon budget after applying each calibrated $f_{\rm esc}^{\rm gal}$ prescription. Dashed curves show the comparison runs where the additional AGN contribution or CGM attenuation is switched off. The shaded region shows the 2.5--97.5\% of the Constant-$f_{\rm esc}$ posterior from \citetalias{Qin2025PASA...42...49Q}. The weighting generally shifts the ionizing photon budget towards fainter galaxies, although the strength of this shift varies between models. Approximate observational limits at $z\sim6$ for various deep imaging surveys are indicated, which only probe the bright end of the ionizing sources.}
	\label{fig:fion_muv}
\end{figure}

Finally, Fig.~\ref{fig:fion_muv} summarizes these source-weighting differences in terms of the observable galaxy population. The figure shows the cumulative fraction of ionizing photons contributed by galaxies brighter than a given UV magnitude at $z=6$. It shows that roughly half of the intrinsic ionizing photon production comes from galaxies brighter than $M_{\rm UV}\sim {-}15$ to ${-}16$, with a substantial remaining contribution from fainter systems. This is below the luminosity range most robustly constrained by deep surveys, which typically reach $M_{\rm UV}\gtrsim -17$ at high redshift, although lensing studies can probe fainter magnitudes with larger systematic uncertainties.

Applying $f_{\rm esc}^{\rm gal}$ shifts the escaping photon budget to fainter galaxies in most of the calibrated models. The fiducial, AGN-assisted, CGM-attenuated, stellar-mass-dependent, and SFR-dependent models are closely grouped, with $\sim50$ per cent of the escaping photons coming from galaxies fainter than $M_{\rm UV}{\sim} {-}13$ to ${-}14$. The SF-disc model remain closer to the intrinsic distribution, consistent with their flatter effective $f_{\rm esc}^{\rm gal}$--$M_{\rm vir}$ relations and their comparatively larger contribution from more massive, brighter galaxies. The sSFR-dependent model lies between these cases, because its escape-fraction scaling mainly changes the normalization with redshift rather than imposing a strong monotonic dependence on halo mass.

The comparison with the Constant-$f_{\rm esc}$ model from \citetalias{Qin2025PASA...42...49Q} is encouraging. Despite using a different galaxy--halo connection and a more detailed galaxy-formation model, our calibrated models occupy the same broad region of source-luminosity space. This reinforces the main conclusion of this section: the Ly$\alpha$ forest calibration allows some freedom in how ionizing photons are redistributed among halo masses and galaxy luminosities, but it consistently favours scenarios in which faint galaxies provide the majority of the escaping ionizing photon budget.

\section{Robustness of Ly$\alpha$ forest-inferred reionization histories}\label{sec:robust_eor}
The goal of this section is to test the robustness of Ly$\alpha$ forest-inferred reionization histories to changes in the ionizing source model. Having shown in Sec.~\ref{sec:fesc} that different escape-fraction prescriptions can redistribute the escaping ionizing photon budget across halo mass and UV luminosity, we now ask whether these differences lead to substantially different reionization histories once each model is calibrated to the same Ly$\alpha$ forest data.

\begin{figure*}
	\centering
	\includegraphics[width=\textwidth]{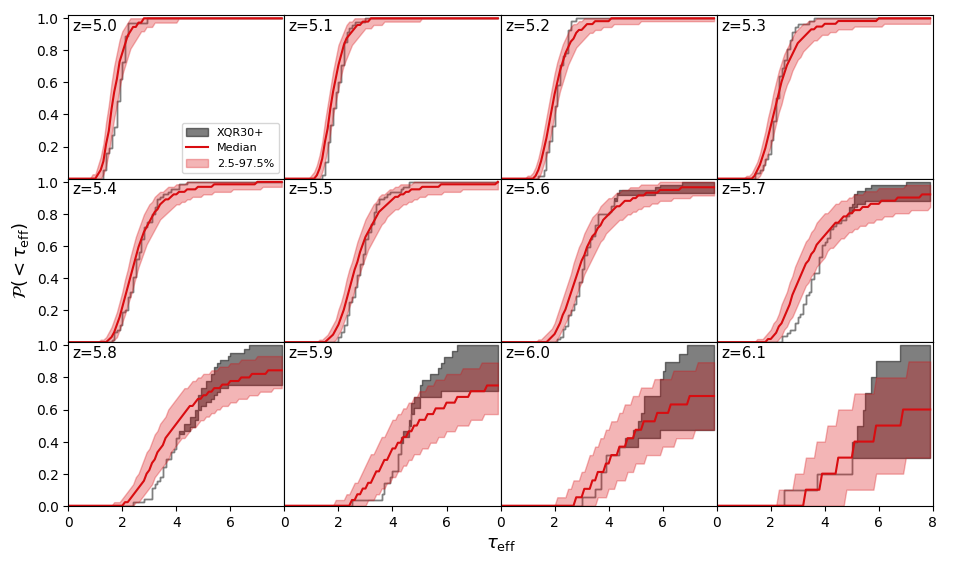}
	\caption{Cumulative distribution functions of the Ly$\alpha$ effective optical depth, $\tau_{\rm eff}$, for our fiducial model over $5.0 \leq z \leq 6.1$. The model CDFs are constructed from mock samples with the same number of sightlines as the XQR-30+ dataset to reflect the expected sample variance. The solid curve shows the median CDF, and the coloured shaded region shows the 2.5th--97.5th percentile range across mock samples. The observed XQR-30+ CDFs are shown in grey, and for bins with non-detections, the shaded region indicates the range obtained by assigning the transmitted flux to either zero or twice the noise level \citep{Bosman2022MNRAS.514...55B}.}	
	\label{fig:cdf_fiducial}
\end{figure*}

We first examine the fiducial halo-mass-dependent escape fraction model, which provides the baseline for the rest of this work. Fig.~\ref{fig:cdf_fiducial} compares the cumulative distribution functions (CDFs) of the Ly$\alpha$ effective optical depth, $\tau_{\rm eff}$, predicted by this model with the XQR-30+ measurements \citep{Bosman2022MNRAS.514...55B}. Overall, the fiducial model is well calibrated to the data across the redshift range considered with minor differences visibly at $z\sim5.7$--5.8. It reproduces both the rapid evolution of the opacity distribution with redshift and the broadening of the CDF towards higher optical depths at $z\gtrsim5.6$, where the forest becomes increasingly sensitive to spatial fluctuations in the ionizing background and residual neutral gas.

The corresponding global reionization quantities are shown in Fig.~\ref{fig:xH_mvir}. The fiducial model predicts a clumping factor that increases towards lower redshifts, as progressively denser and more structured regions of the IGM become ionized and contribute to the recombination budget \citep{Furlanetto2006PhR...433..181F,Finlator2012MNRAS.427.2464F}. Its evolution is broadly consistent with the effective clumping-factor prescription proposed in \citetalias{Qin2025PASA...42...49Q}, starting close to unity at early times, increasing by roughly a factor of two by the EoR midpoint, and reaching $C\sim10$ by the end of reionization. The resulting global neutral fraction history, $x_{\rm HI}$, is also consistent with the late and extended EoR inferred by \citetalias{Qin2025PASA...42...49Q},\footnote{We show the Constant-$f_{\rm esc}$ model from \citetalias{Qin2025PASA...42...49Q}, which includes numerical corrections to conserve ionizing photons and therefore also contributes to producing slightly earlier ionization \citep{Park2022MNRAS.517..192P}.} as well as with most observational constraints from measurements of dark pixels, damping wings, and Ly$\alpha$ emitters (references listed in the caption). This EoR history leads to a CMB optical depth of $\tau_e=0.0636$, consistent with \citet{Planck2020A&A...641A...6P} and the reanalysis by \citet{Qin2020MNRAS.499..550Q} with $\tau_e=0.0569^{+0.0081}_{-0.0066}$. Similarly, the evolution of the comoving ionizing emissivity, $\dot{n}_{\rm ion}$, lies close to previous estimates over the redshift range relevant for the Ly$\alpha$ forest. A sharp late-time drop in emissivity, as seen in several earlier simulations (e.g. \citealt{Kulkarni2019MNRAS.485L..24K,Keating2020MNRAS.491.1736K,Ocvirk2021MNRAS.507.6108O}), is not required in our model. Instead, we find a more gradual plateau, in which the relatively high emissivity is balanced by enhanced small-scale absorption associated with the increasing clumping factor. This agreement provides a useful baseline: despite using a more detailed galaxy-formation model than \citetalias{Qin2025PASA...42...49Q} and a different treatment of the galaxy--halo connection, the fiducial model recovers a similar broad reionization history when calibrated to the current Ly$\alpha$ forest data.

\subsection{Late-time source modifications: AGN \& CGM attenuation}\label{subsec:agn_cgm_results}

\begin{figure*}
	\centering
	\includegraphics[width=0.503\textwidth]{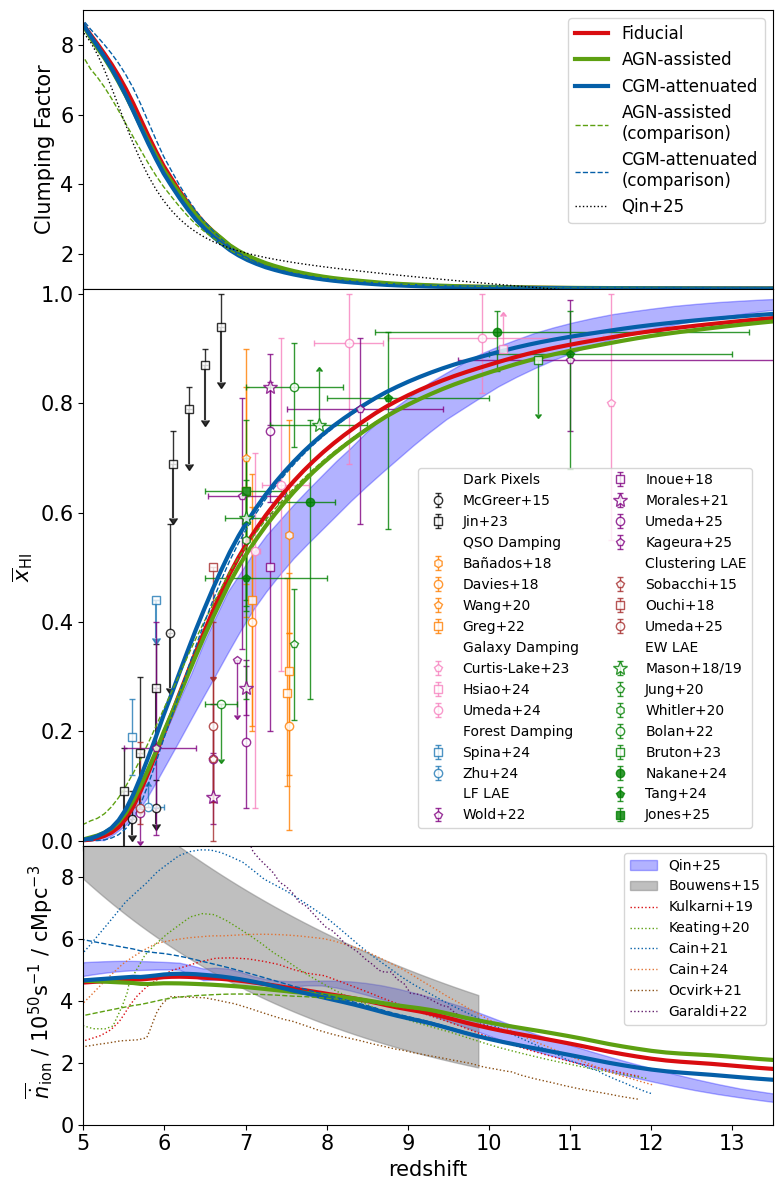}\hfill
	\includegraphics[width=0.497\textwidth]{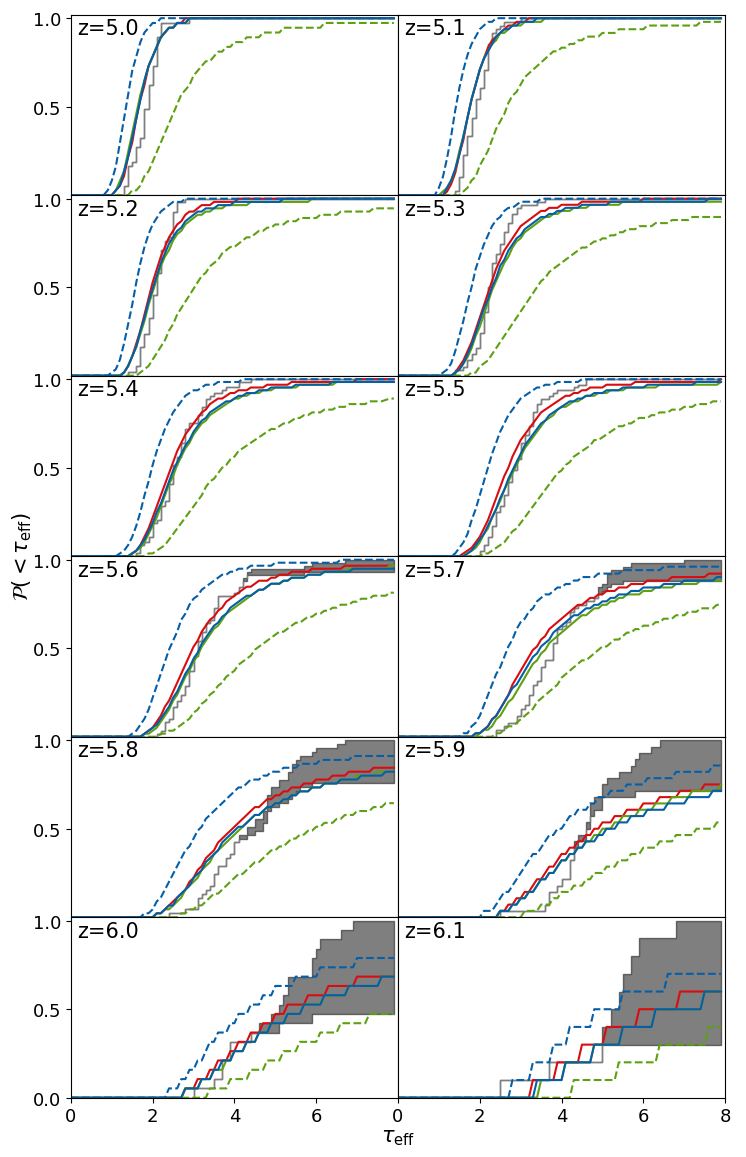}
	\caption{\textit{Left-hand panels}: reionization histories and ionizing emissivity evolution for the fiducial, AGN-assisted, and CGM-attenuated models. While solid curves show the three calibrated models used in this work, the corresponding dashed curves show comparison runs in which the additional AGN contribution or CGM attenuation is switched off. Note that the two comparison runs are largely overlapped with their reference ones at $z\gtrsim7$.		
		The top panel shows the evolution of the volume-averaged clumping factor, $C$, in comparison with the effective clumping factor proposed by \citetalias{Qin2025PASA...42...49Q}. The middle panel shows the volume-averaged neutral fraction, $\overline{x}_{\rm HI}$, compared with \citetalias{Qin2025PASA...42...49Q} (2.5--97.5\% of the Constant-$f_{\rm esc}$ posterior) and other observational constraints from measurements of dark pixels, damping wings and Ly$\alpha$ emitters \citep{McGreer2015MNRAS.447..499M,Jin2023ApJ...942...59J,Banados2018Natur.553..473B,Davies2018ApJ...864..142D,Wang2020ApJ...896...23W,Greig2022,Curtis-Lake2023NatAs...7..622C,Hsiao2024ApJ...973....8H,Umeda2024ApJ...971..124U,Spina2024A&A...688L..26S,Zhu2024MNRAS.533L..49Z,Wold2022ApJ...927...36W,Inoue2018PASJ...70...55I,Morales2021ApJ...919..120M,Umeda2025ApJS..277...37U,Kageura2025ApJS..278...33K,Sobacchi2015MNRAS.453.1843S,Ouchi2018PASJ...70S..13O,Mason2018ApJ...856....2M,Mason2019MNRAS.485.3947M,Jung2020ApJ...904..144J,Whitler2020MNRAS.495.3602W,Bolan2022MNRAS.517.3263B,Bruton2023ApJ...949L..40B,Nakane2024ApJ...967...28N,Bolan2022MNRAS.517.3263B,Bruton2023ApJ...949L..40B,Nakane2024ApJ...967...28N,Tang2024ApJ...975..208T,Jones2025MNRAS.536.2355J}. The bottom panel shows the comoving ionizing emissivity, $\dot{n}_{\rm ion}$, compared with other simulation results (\citetalias{Qin2025PASA...42...49Q}, \citealt{Kulkarni2019MNRAS.485L..24K,Keating2020MNRAS.491.1736K,Cain2021ApJ...917L..37C,Cain2024MNRAS.531.1951C,Ocvirk2021MNRAS.507.6108O,Garaldi2022MNRAS.512.4909G}) and an empirical relation from the observed SFRD \citep{Bouwens2015b}. \textit{Right-hand panels}: the median Ly$\alpha$ CDFs from the various models. Note that their scatters due to sample variance are similar to the fiducial model shown in Fig. \ref{fig:cdf_fiducial}.}
	\label{fig:xH_mvir}
\end{figure*}

Fig.~\ref{fig:xH_mvir} also compares the fiducial model with two extensions that modify the source-side ionizing emissivity: the AGN-assisted model and the CGM-attenuated model. For each extension, we also show a comparison run in which the additional effect is switched off while keeping the calibrated $f_{\rm esc}^{\rm gal}$ prescription fixed. These comparison runs are useful for isolating the effect of the extra AGN/CGM term from the change in the underlying stellar escape fraction parameters.

The AGN-assisted model first differs from the fiducial model through its stellar component. Its calibrated stellar escape fraction prescription has a lower normalization and a steeper halo-mass dependence, $(f_0,\alpha)=(0.06,-0.4)$, compared with $(0.08,-0.3)$ in the fiducial model. This shifts the escaping stellar emissivity further towards low-mass halos. At early times, when low-mass galaxies dominate the available source population, this allows reionization to start slightly earlier than in the fiducial model. At later times, however, the same prescription reduces the relative contribution from more massive galaxies, which grow rapidly towards the end of reionization. The stellar component alone therefore produces too little late-time ionizing emissivity. This is shown by the AGN-assisted comparison run in Fig.~\ref{fig:xH_mvir}, where the AGN contribution is switched off while the same stellar escape fraction prescription is retained. We see that the late-stage ionization history is slower and the Ly$\alpha$ CDFs show systematically higher opacities, especially at $z\lesssim5.8$.

In the full AGN-assisted model, the missing late-time stellar emissivity is partly compensated by ionizing photons from accreting blackholes. The AGN contribution rises rapidly towards lower redshift as the blackhole population grows, bringing both the late-time ionization history and the Ly$\alpha$ CDFs back close to the fiducial model and the observations, with a resulting CMB optical depth of $\tau_e=0.0652$. On average\footnote{This is also reflected in the morphology of the ionization field that we do not find large-scale regions whose ionization is driven primarily by AGN. This is expected because the most massive blackholes reside in massive galaxies. AGN therefore boost the emissivity in already source-rich environments, rather than generating a distinct population of AGN-dominated ionized bubbles (see the emissivty panel in Fig.~\ref{fig:lc_qso}).}, the AGN contribution is still subdominant in this calibrated model, reaching at most $\sim15$ per cent of the stellar ionizing emissivity at $z\sim5$, which is why we refer to this model as AGN-assisted rather than AGN-dominated. The resulting picture is therefore that low-mass galaxies allow a slightly earlier onset of reionization, while a modest AGN contribution helps maintain the ionizing emissivity required near the end of reionization. We note that the exact level of AGN contribution remains uncertain, which depends on the abundance of faint AGN, their duty cycles, spectral hardening, and ionizing escape fractions (see footnote \ref{footnote:fescAGN}). A larger AGN contribution could in principle allow even more early contribution from low-mass galaxies while still matching the forest at late times. Within the level of AGN contribution explored here, however, this extra freedom is limited.

The CGM-attenuated model explores the opposite direction. Its intrinsic stellar escape fraction scaling, $(f_0,\alpha)=(0.1,-0.2)$, is shallower than in the fiducial model, allowing a relatively larger contribution from massive halos before attenuation is applied. The comparison run without CGM attenuation shows what this intrinsic source model would do on its own: as massive galaxies become increasingly important towards lower redshifts, the escaping stellar emissivity remains too high, producing too much late-time Ly$\alpha$ transmission and ionizing the IGM too early. This earlier ionization also exposes progressively denser and more structured gas to the ionizing background, leading to a faster rise in the clumping factor and hence in the recombination budget.

\begin{figure*}
	\centering
	\includegraphics[width=0.503\textwidth]{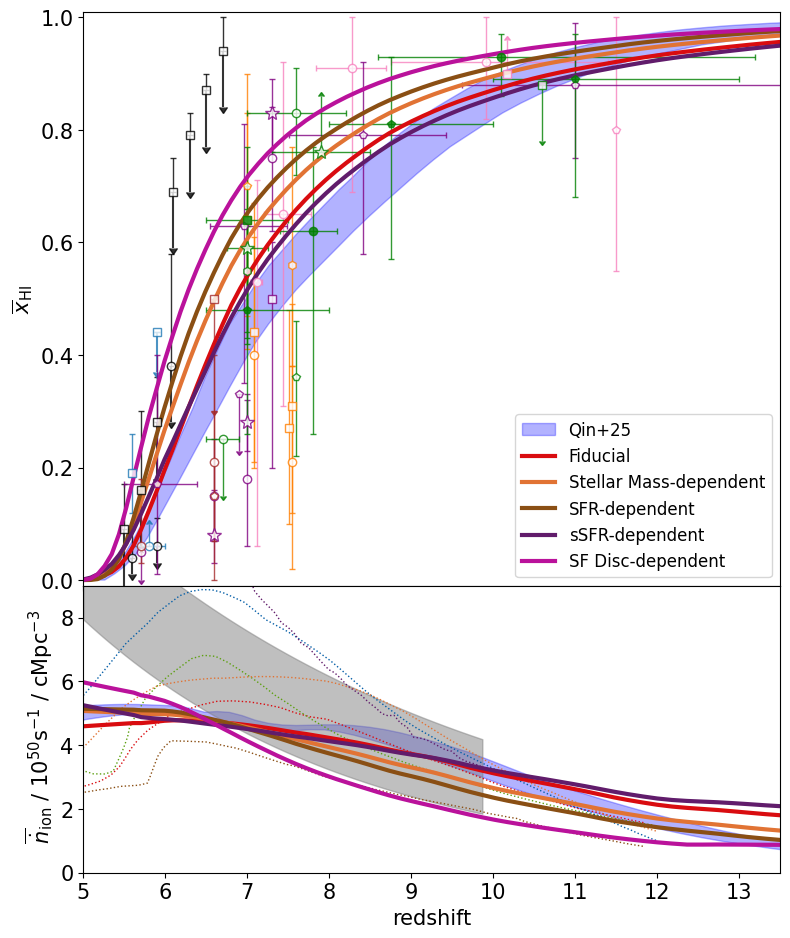}\hfill
	\includegraphics[width=0.497\textwidth]{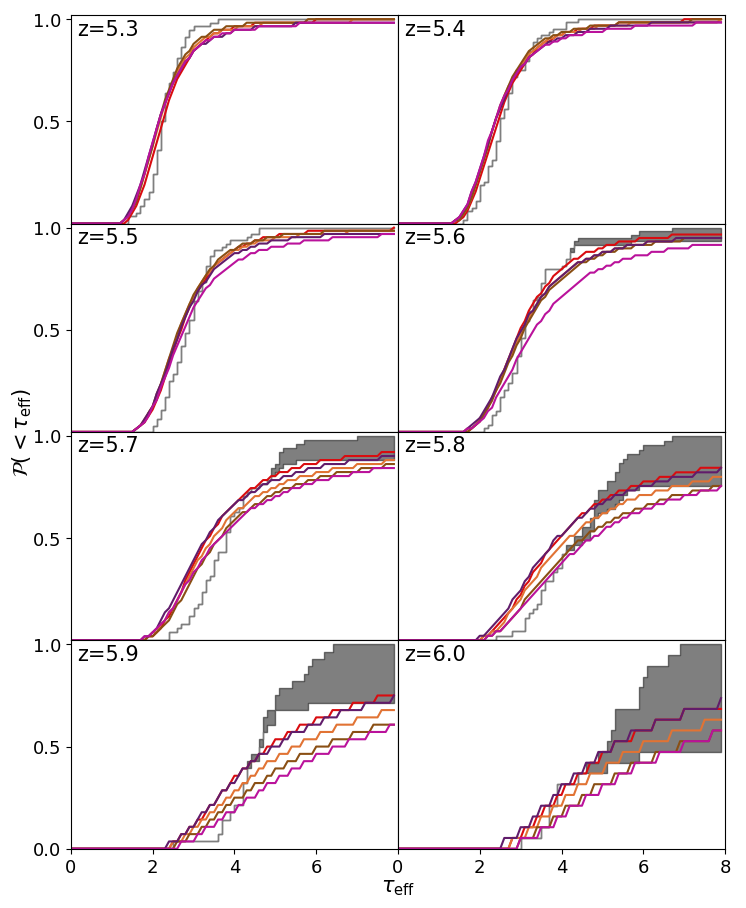}
	\caption{Similar to Fig. \ref{fig:xH_mvir} but for galaxy-property-dependent models and with fewer panels for clarity.}
	\label{fig:xH_galp}
\end{figure*}

In the full CGM-attenuated model, the additional attenuation suppresses the escaping stellar emissivity towards the end of reionization by up to $\sim15$ per cent. It brings the Ly$\alpha$ CDFs and the global ionization history back into closer agreement with the fiducial model and the observations, with a resulting CMB optical depth of $\tau_e=0.0609$. We interpret this attenuation as an effective CGM opacity, such that $f_{\rm esc}^{\rm gal}$ is further reduced by a factor $\exp(-\tau_{\rm CGM})$. Here, $\tau_{\rm CGM}$ represents unresolved absorption by dense gas structures around galaxies and the response of this absorption to gas structure is expected to be non-linear: recombinations scale as density squared, and self-shielding can further increase both the neutral column and the covering fraction of optically thick absorbers once the gas becomes sufficiently dense. We therefore adopt a steep $\tau_{\rm CGM}\propto C^3$ scaling as a phenomenological prescription for unresolved CGM absorption. Since the clumping factor remains modest at early times but rises rapidly towards the end of reionization, this term has little impact on the early emissivity while providing the late-time suppression (see the attenuated $f_{\rm esc}^{\rm gal}$ in the middle leftmost panel of Fig.~\ref{fig:fesc_gal}) needed to match the Ly$\alpha$ forest CDFs. But we also caution interpreting the exponent as a first-principle prediction of the CGM opacity--clumping relation: it depends on unresolved properties of high-redshift circumgalactic gas, including the abundance, geometry, and ionization state of small-scale absorbers. Stronger attenuation would allow a more massive-halo-weighted intrinsic source model, while weaker attenuation would move the calibrated solution back towards the fiducial case.

Taken together, the AGN-assisted and CGM-attenuated models move the source model in opposite directions. The AGN-assisted model adds late-time emissivity to a more low-mass-weighted stellar population, while the CGM-attenuated model removes late-time escaping photons from a more massive-halo-weighted stellar population. Both effects broaden the range of viable source histories, allowing modest shifts in the early stages of reionization. However, after calibration to the Ly$\alpha$ forest, both models still converge towards a similar late-time ionization history. We therefore conclude that the inferred EoR history is robust: while the detailed allocation of ionizing photons among low-mass galaxies, massive galaxies, AGN, and CGM absorption is model-dependent, the Ly$\alpha$ forest strongly restricts the final stages of reionization. Moreover, if the effective halo-mass dependence of the ionizing emissivity remains approximately fixed towards higher redshift (as assumed in this subsection), the earlier evolution of the neutral fraction is also tightly constrained. In the next subsection, we relax this last assumption by tying $f_{\rm esc}^{\rm gal}$ directly to evolving galaxy properties rather than to halo mass alone.

\subsection{Source stochasticity: galaxy-property-dependent escape fractions}\label{subsec:galaxy_property_fesc}

The previous subsection explored source-side modifications that act on top of a halo-mass-dependent stellar escape fraction. We now ask whether the inferred EoR history changes when the effective dependence of $f_{\rm esc}^{\rm gal}$ on halo mass is allowed to evolve indirectly through the galaxy population itself. This is motivated by the fact that galaxy properties such as stellar mass, star-formation rate, specific star-formation rate, and star-forming disc surface density do not map onto halo mass in a fixed way. Their relations with $M_{\rm vir}$ evolve with redshift and have substantial scatter at fixed halo mass, as shown in the upper panels of Fig.~\ref{fig:fesc_gal}. A power-law escape fraction in one of these quantities can therefore produce an effective $f_{\rm esc}^{\rm gal}$--$M_{\rm vir}$ relation that changes over the duration of reionization.

Fig.~\ref{fig:xH_galp} compares the median Ly$\alpha$ effective optical depth CDFs for the galaxy-property-dependent models against the fiducial and the XQR-30+ measurements. By construction, all of these models have been calibrated to the same Ly$\alpha$ forest data, and they therefore reproduce the observed CDFs at a broadly similar level. The differences between models are modest at $z\lesssim5.6$, where the forest is already mostly transmitting in the lower-opacity regions. At higher redshifts, especially around $z\sim5.7$--$6.0$, the model-to-model differences become more visible because the CDFs are more sensitive to the abundance of long, high-opacity segments. Even there, however, the calibrated models remain within a relatively narrow range compared to the full evolution of the observed opacity distribution.

The corresponding neutral fraction histories and ionizing emissivities are also shown in Fig.~\ref{fig:xH_galp}. The stellar-mass- and SFR-dependent models remain close to the fiducial model including their CMB optical depths ($\tau_e = 0.0596$ and 0.0571). This is not surprising, because both $M_*$ and SFR are strongly correlated with $M_{\rm vir}$ over the halo-mass range that dominates the ionizing photon budget (see the second and third columns of Fig.~\ref{fig:fesc_gal}). Their escape-fraction prescriptions therefore behave like slightly scattered versions of a halo-mass-dependent model, leading to similar emissivity evolution and similar reionization histories. 

The sSFR-dependent model introduces more scatter at fixed halo mass (see the fourth column of Fig.~\ref{fig:fesc_gal}), but its global reionization history (with $\tau_e=0.0652$) is still close to the fiducial case. This is because the positive scaling with sSFR maps, on average, into a higher escape fraction for lower-mass galaxies, broadly preserving the same qualitative source hierarchy as the fiducial model. In other words, the sSFR model adds stochasticity to the galaxy emissivity field, but does not strongly change which part of the halo population dominates the photon budget.

The star-forming-disc model shows the largest departure among the galaxy-property prescriptions. After calibration, its effective $f_{\rm esc}^{\rm gal}$--$M_{\rm vir}$ relation is nearly flat over the halo-mass range that dominates the ionizing photon budget. As a result, the escape-fraction weighting does not strongly shift photons towards low-mass halos, unlike the other models. This makes the SF-disc model the most massive-halo-weighted scenario in our suite. Despite this different source weighting, the model is still calibrated to the Ly$\alpha$ forest and approaches a similar late-time ionization history with $\tau_e=0.0535$. 

These galaxy-property-dependent models therefore provide a useful test of source stochasticity. Tying $f_{\rm esc}^{\rm gal}$ to evolving galaxy properties can broaden the allowed source histories more than the purely halo-mass-dependent model, especially at earlier times. However, the Ly$\alpha$ forest calibration still pulls the models towards a similar late-time reionization history. The main conclusion from this comparison is therefore consistent with the AGN and CGM tests above: additional freedom in the source model can change how ionizing photons are assigned to galaxies and can modestly shift the early evolution of $\overline{x}_{\rm HI}$, but the late stages of reionization remain tightly restricted by the Ly$\alpha$ forest opacity distribution.

\section{Summary discussion}\label{sec:discussion}

In this work, we revisited Ly$\alpha$ forest constraints on the EoR using \textsc{Meraxes}, a semi-analytic model of galaxy formation and reionization. The goal was to test whether the tight reionization histories inferred by Ly$\alpha$ forest forward modelling, $z|_{\overline{x}_{\rm HI}=0.5}=7.7{\pm}0.1$ and $z|_{\overline{x}_{\rm HI}=0.05}=5.44{\pm}0.02$ (\citetalias{Qin2025PASA...42...49Q}), remain robust once the ionizing source population is allowed to be more flexible than in simple halo-based prescriptions. 
We constructed a suite of models in which the galaxy escape fraction is tied either to halo mass or to galaxy properties, including stellar mass, SFR, sSFR, and star-forming disc surface density. We also explored two extensions to the fiducial halo-mass-dependent model: an AGN-assisted model and a CGM-attenuated model. All models were calibrated against the observed Ly$\alpha$ effective-optical-depth CDFs from XQR-30+ \citep{Bosman2022MNRAS.514...55B}. Despite their different source prescriptions, the calibrated models predict reionization milestones over a relatively narrow range, with the midpoint occurring between z$\sim6.2$ and 6.9 and the late stages at $z\sim5.36$--5.57.

The fiducial \textsc{Meraxes} model, in which $f_{\rm esc}^{\rm gal}$ increases towards lower halo masses, reproduces the observed Ly$\alpha$ forest opacity distributions across $5\lesssim z\lesssim6.1$ at a level comparable to previous forward-modelling analyses. The corresponding reionization history is late and extended, and is broadly consistent with \citetalias{Qin2025PASA...42...49Q}. The fiducial model also predicts a clumping factor that rises towards lower redshifts, reaching values of order $C\sim10$ by the end of reionization. In this model, a relatively high ionizing emissivity can be maintained without over-ionizing the IGM because the increasing small-scale clumping boosts the recombination and absorption budget.

Changing the escape-fraction prescription redistributes the escaping ionizing photon budget across the galaxy population. The intrinsic ionizing photon production is nearly the same across our model suite because the underlying \textsc{Meraxes} galaxy catalogue is fixed after being calibrated against the observed UV luminosity functions at high redshift. Its redshift evolution is driven mainly by galaxy formation: towards lower redshift, more massive halos contribute an increasing fraction of the intrinsic photon production. Applying $f_{\rm esc}^{\rm gal}$ generally shifts the escaping photon budget back towards lower-mass halos and fainter galaxies. However, the strength of this shift depends on the source model. The fiducial, AGN-assisted, CGM-attenuated, stellar-mass-dependent, and SFR-dependent models produce relatively similar low-mass weighting, while the sSFR-dependent model is flatter with halo mass and the SF-disc model remains more weighted towards massive halos.

The AGN-assisted model shows that modest AGN contributions can provide additional freedom without making reionization AGN-dominated. In our calibrated model, the stellar component is weighted more strongly towards low-mass halos than in the fiducial case, which allows reionization to start slightly earlier but reduces the late-time stellar emissivity. Accreting blackholes then supply a late-time emissivity boost as the blackhole population grows. The AGN contribution remains subdominant, reaching at most $\sim15$ per cent of the stellar ionizing emissivity at $z\sim5$, but is sufficient to bring the late-time Ly$\alpha$ CDFs and ionization history back close to the fiducial model. Larger or more uncertain AGN contributions could allow somewhat more freedom in the early source history, but within the range explored here the effect on the recovered EoR history is modest.

The CGM-attenuated model provides a complementary test in the opposite direction. By adopting a shallower intrinsic halo-mass dependence, this model allows a larger relative contribution from massive halos before attenuation is applied. Without CGM attenuation, it produces too much late-time Ly$\alpha$ transmission and ionizes the IGM too early. A phenomenological CGM opacity term, increasing steeply with the local clumping factor, suppresses the escaping stellar emissivity near the end of reionization and restores agreement with the Ly$\alpha$ forest measurements. The required clumping dependence is an effective way to capture unresolved dense absorbers around galaxies. Varying the strength of this attenuation would move the calibrated solution between a more massive-halo-weighted model and the fiducial one that is more low-mass dominated.

Galaxy-property-dependent escape fractions introduce source stochasticity but do not strongly alter the late-time reionization history once calibrated to the forest data. The stellar-mass- and SFR-dependent models closely resemble the fiducial model because both $M_*$ and SFR are strongly correlated with halo mass. The sSFR-dependent model introduces more scatter and a flatter effective halo-mass dependence, while the SF-disc model is the most massive-halo-weighted case and requires additional CGM suppression to match the Ly$\alpha$ forest observation. Despite these differences, the calibrated models produce similar Ly$\alpha$ CDFs and converge towards similar late-time ionization histories.
	
These results clarify the nature of the Ly$\alpha$ forest constraint. The forest strongly restricts the timing and progression of the final stages of reionization because the opacity distribution is highly sensitive to residual neutral gas and spatial fluctuations in the ionizing background. However, the Ly$\alpha$ CDFs alone do not uniquely determine how the required ionizing photons are allocated among low-mass galaxies, massive galaxies, AGN, and unresolved absorbers. The robust inference is therefore the late and extended reionization history, while the detailed physical origin of the ionizing photon budget remains model-dependent. In this sense, the spread among our calibrated \textsc{Meraxes} models provides an estimate of the source-modelling systematic uncertainty on Ly$\alpha$ forest-based EoR inference.
	
\section{Conclusions}\label{sec:conclusion}

The main result of this work is that Ly$\alpha$ forest-inferred reionization histories are robust across a broad suite of galaxy-formation-based source models. Although varying the escape-fraction prescription changes the allocation of escaping ionizing photons among low-mass galaxies, massive galaxies, AGN, and unresolved CGM absorption, all viable models calibrated to the XQR-30+ Ly$\alpha$ forest opacity distributions converge towards a similar late and extended EoR. Across the calibrated models explored here, the midpoint of reionization lies in the range of $z|_{\overline{x}_{\rm HI}=0.5}\sim6.2$--6.9, while the final stages occur at $z|_{\overline{x}_{\rm HI}=0.05}\sim5.36$--5.57. This limited model-to-model spread indicates that the high-redshift Ly$\alpha$ forest strongly constrains the final stages of reionization, even when the ionizing source population is allowed to vary substantially within physically motivated galaxy-formation models.

\section*{Data Availability}

The data underlying this article will be shared on reasonable request to the corresponding author. 

\section*{Acknowledgements}
The authors gratefully acknowledge the HPC support from OzSTAR and Gadi in Australia. YQ is supported by the ARC Discovery Early Career Researcher Award (DECRA) through fellowship \#DE240101129.




\bibliographystyle{mnras}
\bibliography{references} 




\appendix

\label{lastpage}
\end{document}